\pdfoutput=1
\documentclass[preprint,12pt]{elsarticle}

\usepackage[utf8]{inputenc}
\usepackage{amsmath,amssymb,amsthm}
\usepackage{thmtools}
\usepackage{mathtools}
\usepackage{enumitem}
\usepackage{xcolor}
\usepackage{graphicx}
\usepackage{tikz}
\usepackage{tikzit}
\usepackage{tkz-graph}
\usepackage{caption}
\usepackage{subcaption}
\usepackage{algorithm}
\usepackage{algpseudocode}
\algrenewcommand\algorithmicrequire{\textbf{Input:}}
\algrenewcommand\algorithmicensure{\textbf{Output:}}
\algblockdefx[Foreach]{ForEach}{EndFor}[1]{\textbf{for each} #1 \textbf{do}}{\textbf{end for}}
\usepackage[hypertexnames=false]{hyperref}
\usepackage[capitalise]{cleveref}
\crefname{observation}{Observation}{Observations}
\crefname{condition}{condition}{conditions}
\Crefname{condition}{Condition}{Conditions}
\crefname{claim}{Claim}{Claims}
\Crefname{claim}{Claim}{Claims}
\crefname{open}{Open Problem}{Open Problems}
\Crefname{open}{Open Problem}{Open Problems}

\tikzstyle{vertex}=[circle, fill=black, draw, inner sep=0pt, minimum size=5pt]
\newcommand{\vertex}{\node[vertex]}

\definecolor{palEstructura}{HTML}{B33149}
\definecolor{palTheorem}{HTML}{E87474}
\definecolor{palLemma}{HTML}{52AFE8}
\definecolor{palCorollary}{HTML}{3B54B0}
\definecolor{palDefinition}{HTML}{C6DE3C}
\definecolor{palObservation}{HTML}{52E3C8}
\definecolor{palClaim}{HTML}{F2B03E}
\definecolor{palOpen}{HTML}{B168D6}
\definecolor{palCompA}{HTML}{F09342}
\definecolor{palCompB}{HTML}{1C7FC4}
\definecolor{palCompC}{HTML}{2FB89C}
\definecolor{palCompD}{HTML}{7425A8}
\definecolor{palCompE}{HTML}{DDF05A}
\definecolor{palClassA}{HTML}{E04444}
\definecolor{palClassB}{HTML}{3657CC}
\definecolor{palClassC}{HTML}{B351E8}
\definecolor{palResaltado}{HTML}{F09342}
\definecolor{degClawOne}{HTML}{4E0D9D}
\definecolor{degClawThreeFigA}{HTML}{C54257}
\definecolor{degClawFourFigB}{HTML}{FEF181}
\definecolor{degClawTwoFigB}{HTML}{2CC1AC}
\definecolor{degClawOneFigB}{HTML}{4E0D9D}
\definecolor{degClawFourFigC}{HTML}{FEF181}
\definecolor{degClawThreeFigC}{HTML}{C54257}
\definecolor{degClawTwoFigC}{HTML}{2CC1AC}
\definecolor{degLotFour}{HTML}{0DA78D}
\definecolor{degLotFourFigB}{HTML}{37BDA2}
\definecolor{degLotFourFigC}{HTML}{37BDA2}
\definecolor{degLotTwo}{HTML}{FEF181}
\definecolor{escAmbarD}{HTML}{FFCB59}
\definecolor{escAmbarE}{HTML}{FFCB59}
\definecolor{escAzul}{HTML}{4A9BE3}
\definecolor{escAzulE}{HTML}{4A9BE3}
\definecolor{escBordoD}{HTML}{C50040}
\definecolor{escBordoE}{HTML}{C50040}
\definecolor{escLimaF}{HTML}{FEF181}
\definecolor{escVioletaD}{HTML}{A04ECE}
\definecolor{escVioletaE}{HTML}{A04ECE}
\definecolor{escVioletaF}{HTML}{AE5BDC}
\definecolor{degLotUone}{HTML}{B8B8B8}

\declaretheorem[name=Theorem, numberwithin=section]{theorem}
\declaretheorem[name=Lemma, sibling=theorem]{lemma}

\declaretheorem[name=Proposition, sibling=theorem]{proposition}

\declaretheorem[name=Observation, sibling=theorem]{observation}
\declaretheorem[name=Theorem, numbered=no]{introtheorem}
\declaretheorem[name=Definition, sibling=theorem, style=definition]{definition}

\declaretheorem[name=Open Problem, sibling=theorem, style=definition]{open}

\declaretheorem[name=Claim, sibling=theorem, style=remark]{claim}

\newcommand*{\CompOne}{\ensuremath{Comp_1}}
\newcommand*{\Gplusu}{G^{+u}}
\newcommand*{\lotus}{\textnormal{lotus}}
\newcommand{\thin}{\ensuremath{\operatorname{thin}}}
\newcommand*{\condi}{\hyperlink{sc:cond-i}{(i)}}
\newcommand*{\condii}{\hyperlink{sc:cond-ii}{(ii)}}
\newcommand{\pthin}{\ensuremath{\operatorname{pthin}}}
\newcommand{\lmimw}{\emph{lmimw}}
\newcommand{\pw}{\emph{pw}}
\newcommand{\bw}{\emph{bw}}

\journal{Discrete Applied Mathematics}
\begin{document}

\begin{frontmatter}

\title{Proper thinness at most 2: threshold and trivially perfect graphs}

\author[dc,icc]{B\'arbara Rivero\corref{cor1}}
\ead{brivero@ic.fcen.uba.ar}
\cortext[cor1]{Corresponding author.}
\author[dc,icc]{Eric Brandwein}
\ead{brandweineric@gmail.com}
\author[dc,icc]{Flavia Bonomo-Braberman}
\ead{fbonomo@dc.uba.ar}

\affiliation[dc]{organization={Universidad de Buenos Aires, Facultad de Ciencias Exactas y Naturales, Departamento de Computaci\'on},
            addressline={Pabell\'on Cero + Infinito, Ciudad Universitaria},
            postcode={C1428EGA},
            city={Buenos Aires},
            country={Argentina}}
\affiliation[icc]{organization={CONICET-Universidad de Buenos Aires, Instituto de Investigaci\'on en Ciencias de la Computaci\'on (ICC)},
            addressline={Pabell\'on Cero + Infinito, Ciudad Universitaria},
            postcode={C1428EGA},
            city={Buenos Aires},
            country={Argentina}}

\begin{abstract}
The proper thinness of a graph is a parameter that generalizes proper interval
graphs: the graphs of proper thinness $1$ are exactly the proper interval graphs.
The complexity of computing it is open even for cographs. In this paper we study
it on two subclasses of cographs: threshold graphs and trivially perfect graphs.
For threshold graphs, we show that the proper thinness is $1$ if the graph is
claw-free and $2$ otherwise. For trivially perfect graphs, we characterize those
of proper thinness at most $2$ in three ways: structurally, through two
properties of strongly consistent solutions that we call \emph{staggered} and
\emph{boundary}; algorithmically, with a linear-time algorithm that decides
whether the proper thinness is at most $2$; and by forbidden induced subgraphs,
showing that the \(\mathrm{claw}_2\) and the \lotus{} are exactly the minimal
obstructions.
\end{abstract}


\end{frontmatter}

\section{Introduction}

Many optimization problems defined on graphs are computationally difficult, and it is natural to ask for which graph classes they can be solved efficiently. Knowing that a graph family has \emph{bounded width}, for example bounded treewidth~\cite{R-S-minors2-tw} or clique-width~\cite{Cour-cw}, proved to be an extremely useful property for this purpose: many classical problems, such as graph coloring or maximum independent set, can then be solved efficiently, in many cases through \emph{metatheorems} of very general applicability~\cite{C-M-cw-LinEMSOL,C-M-tree-MSOL}. Other width parameters include pathwidth~\cite{R-S-minors1-pw}, bandwidth~\cite{Pap-bandwidth}, mim-width and its linear version~\cite{VatshelleThesis}, twin-width~\cite{BKTW20twinw}, and thinness~\cite{M-O-R-C-thinness}.

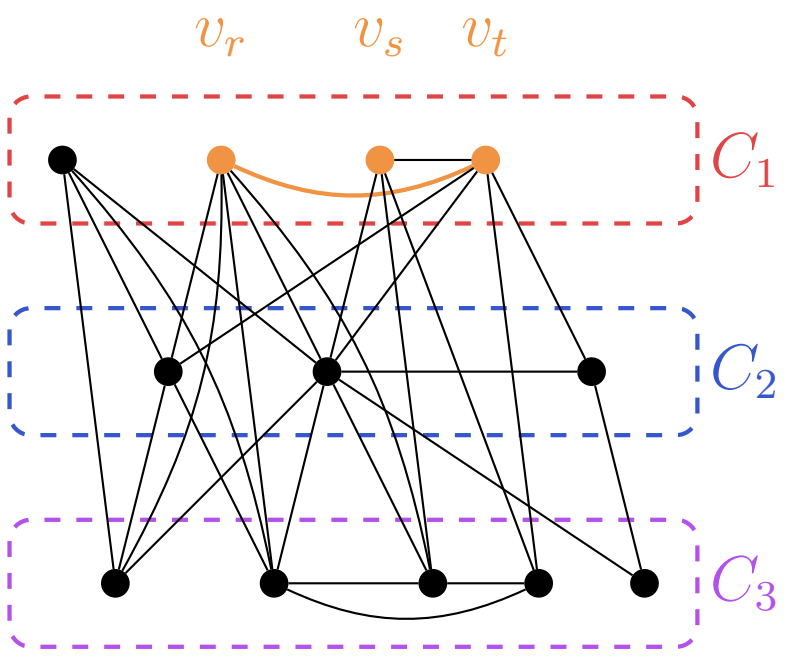
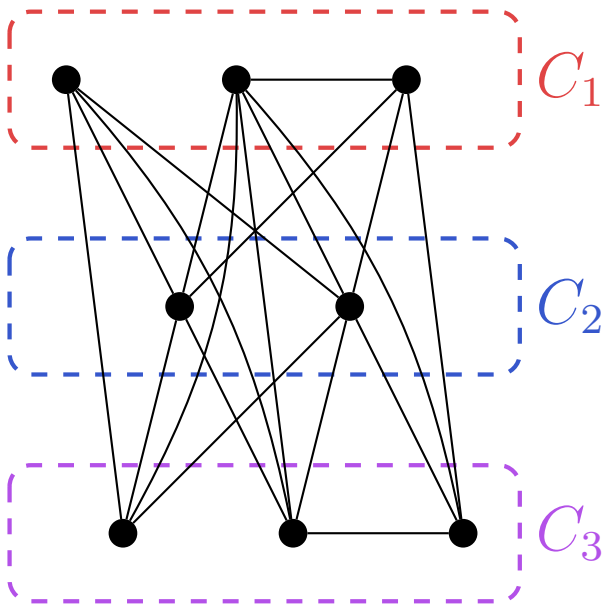
\begin{figure}[ht]
\centering
\begin{subfigure}[b]{0.55\textwidth}
\centering
\begin{tikzpicture}[x={(0cm,-1cm)}, y={(1cm,0cm)}, scale=0.7]
    \vertex(v1) at (-6,4) {};
    \node[vertex, fill=palResaltado, draw=palResaltado] (v2) at (-6,5.5) {};
    \node[vertex, fill=palResaltado, draw=palResaltado] (v3) at (-6,7) {};
    \node[vertex, fill=palResaltado, draw=palResaltado] (v4) at (-6,8) {};

    \vertex(w1) at (-4,5) {};
    \vertex(w2) at (-4,6.5) {};
    \vertex(w3) at (-4,9) {};

    \vertex(z1) at (-2,4.5) {};
    \vertex(z2) at (-2,6) {};
    \vertex(z3) at (-2,7.5) {};
    \vertex(z4) at (-2,8.5) {};
    \vertex(z5) at (-2,9.5) {};

    \node[palResaltado, above, yshift=0.45cm] at (v2.north) {$v_r$};
    \node[palResaltado, above, yshift=0.45cm] at (v3.north) {$v_s$};
    \node[palResaltado, above, yshift=0.45cm] at (v4.north) {$v_t$};

    \draw[dashed, rounded corners, palClassA, thick] (-6.6,3.5) rectangle (-5.4,10);
    \draw[dashed, rounded corners, palClassB, thick] (-4.6,3.5) rectangle (-3.4,10);
    \draw[dashed, rounded corners, palClassC, thick] (-2.6,3.5) rectangle (-1.4,10);
    \node[palClassA] at (-6,10.45) {$C_1$};
    \node[palClassB] at (-4,10.45) {$C_2$};
    \node[palClassC] at (-2,10.45) {$C_3$};

    \draw[palResaltado, thick] (v4) to [bend left=25] (v2);
    \draw (v4) to (v3);

    \draw (w3) to (w2);

    \draw (z4) to [bend left=25] (z2);
    \draw (z4) to (z3);
    \draw (z3) to (z2);

    \draw (v2) to (w1);
    \draw (v3) to (w2);
    \draw (v4) to (w1);
    \draw (v4) to (w2);
    \draw (v1) to (w2);
    \draw (v4) to (w3);
    \draw (v2) to (w2);
    \draw (v1) to (w1);

    \draw (v1) to (z1);
    \draw (v2) to (z2);
    \draw (v2) to [bend left=15] (z1);
    \draw (v1) to [bend left=15] (z2);
    \draw (v2) to [bend left=15] (z3);
    \draw (v4) to (z4);
    \draw (v3) to (z4);
    \draw (v3) to (z3);

    \draw (z2) to (w1);
    \draw (z3) to (w2);
    \draw (z5) to (w3);
    \draw (z5) to (w2);
    \draw (z2) to (w2);
    \draw (z1) to (w1);
    \draw (z1) to (w2);
\end{tikzpicture}
\caption{A consistent solution that is not strongly consistent: in the highlighted triple \(v_r < v_s < v_t\), the vertices \(v_s\) and \(v_t\) are in the same class and \(v_r v_t\) is an edge, but \(v_r v_s\) is not.}\label{fig:thinness}
\end{subfigure}
\hfill
\begin{subfigure}[b]{0.4\textwidth}
\centering
\begin{tikzpicture}[x={(0cm,-1cm)}, y={(1cm,0cm)}, scale=0.75]
    \vertex(v1) at (-6,4) {};
    \vertex(v2) at (-6,5.5) {};
    \vertex(v3) at (-6,7) {};

    \vertex(w1) at (-4,5) {};
    \vertex(w2) at (-4,6.5) {};

    \vertex(z1) at (-2,4.5) {};
    \vertex(z2) at (-2,6) {};
    \vertex(z3) at (-2,7.5) {};

    \draw[dashed, rounded corners, palClassA, thick] (-6.6,3.5) rectangle (-5.4,8);
    \draw[dashed, rounded corners, palClassB, thick] (-4.6,3.5) rectangle (-3.4,8);
    \draw[dashed, rounded corners, palClassC, thick] (-2.6,3.5) rectangle (-1.4,8);
    \node[palClassA] at (-6,8.45) {$C_1$};
    \node[palClassB] at (-4,8.45) {$C_2$};
    \node[palClassC] at (-2,8.45) {$C_3$};

    \draw (v2) to (v3);

    \draw (z3) to (z2);

    \draw (v2) to (w1);
    \draw (v3) to (w2);
    \draw (v1) to (w2);
    \draw (v2) to (w2);
    \draw (v1) to (w1);
    \draw (v3) to (w1);

    \draw (v1) to (z1);
    \draw (v2) to (z2);
    \draw (v2) to [bend left=15] (z1);
    \draw (v1) to [bend left=15] (z2);
    \draw (v2) to [bend left=15] (z3);
    \draw (v3) to (z3);

    \draw (z2) to (w1);
    \draw (z3) to (w2);
    \draw (z2) to (w2);
    \draw (z1) to (w1);
    \draw (z1) to (w2);
\end{tikzpicture}
\caption{A strongly consistent solution: both conditions of \cref{def:proper-thinness} hold for every triple of vertices.}\label{fig:properthinness}
\end{subfigure}
\caption{Two solutions with three classes. The vertices are ordered from left to right, and each class is drawn as a row enclosed in a dashed box.}\label{fig:solutions}
\end{figure}

\textbf{Thinness} was defined by Mannino, Oriolo, Ricci, and Chandran in 2007 in the context of an application to problems arising from cellular telephony related to the independent set problem~\cite{M-O-R-C-thinness}. Informally, a graph is \emph{\(k\)-thin} if its vertices can be ordered and partitioned into \(k\) classes so that, whenever a vertex \(v_t\) is adjacent to a vertex \(v_r\) smaller than \(v_t\), it is also adjacent to every vertex between them that belongs to the class of \(v_r\). The \emph{thinness} \(\thin(G)\) of a graph \(G\) is the smallest such \(k\), and an ordering and a partition with this property are called \emph{consistent} (see \cref{def:thinness} for the formal definition, and \cref{fig:thinness} for an example).

The definition of thinness gave rise to several variants~\cite{B-D-thinness,BGOSS-thin-oper,BOSS-thin-prec-dam-lagos,Mixed-thin-arxiv}. Among them, \emph{proper thinness} was introduced in~\cite{B-D-thinness}, where it is shown that a broad family of problems, including domination problems, can be solved in polynomial time on graphs of bounded thinness or of bounded proper thinness. Informally, a graph is \emph{proper \(k\)-thin} if, in addition, the same holds for the reverse ordering: whenever a vertex \(v_r\) is adjacent to a vertex \(v_t\) greater than \(v_r\), it is also adjacent to every vertex between them that belongs to the class of \(v_t\). The \emph{proper thinness} \(\pthin(G)\) of a graph \(G\) is the smallest such \(k\), and an ordering and a partition with this property are called \emph{strongly consistent} (see \cref{def:proper-thinness} for the formal definition, and \cref{fig:properthinness} for an example).

Clearly \(\thin(G) \leq \pthin(G)\) for every graph \(G\). The difficulty of the proper variant lies in the extra requirement: a strongly consistent solution must be consistent with respect to the ordering \emph{and} to its reverse, so the arguments developed for thinness do not transfer directly. Both parameters are related to other width parameters: \(\lmimw(G) \leq \thin(G) \leq \pw(G)+1\)~\cite{B-D-thinness,M-O-R-C-thinness}, and \(\pthin(G) \leq \max\{1, \bw(G)\}\)~\cite{B-B-lagos21-dam}.

Given an ordering of the vertices, a (strongly) consistent partition into a minimum number of classes can be found in cubic time, whereas, given a partition, it is NP-complete to decide whether there exists a (strongly) consistent ordering for it~\cite{B-D-thinness}. By reducing from that problem, \citet{Shitov-thin} showed that deciding whether a graph is \(k\)-thin is NP-complete. A recent manuscript~\cite{AI-2thin}, generated by GPT-6 Astra Pro prompted by Shitov, shows that this problem remains NP-complete for every fixed \(k \ge 2\), and that computing the proper thinness is NP-hard, even for line graphs of bipartite graphs. The complexity of recognizing proper \(k\)-thin graphs for fixed \(k\) remains open. Computing the thinness is polynomial for cographs~\cite{B-D-thinness} and for trees~\cite{Agus-Eric-DAM25}, and the thinness of cographs has been characterized by forbidden induced subgraphs~\cite{B-D-thinness}, and that of trees structurally~\cite{Agus-Eric-tesis,Agus-Eric-DAM25}. For the proper thinness, both questions remain open in both classes; for trees, only those of proper thinness at most~\(2\) have been characterized~\cite{B-M-P-trees-pthin2}.

In this paper we study the proper thinness of two classical subclasses of cographs: threshold graphs and trivially perfect graphs. We settle the question completely for the former, and up to the value~\(2\) for the latter.

\emph{Threshold graphs}~\cite{Ch-H-thres} are the graphs that can be built from a single vertex by repeatedly adding an isolated or a universal vertex (\cref{def:threshold}), and they are exactly the graphs containing no induced \(P_4\), \(C_4\), or \(2K_2\)~\cite{Ch-H-thres,Ch-H-aggregation}. \emph{Trivially perfect graphs} were characterized by \citet{Wolk1962} as the comparability graphs of rooted forests, and were named by \citet{Golumbic1978}. They are also known as \emph{quasi-threshold graphs}~\cite{YanChenChang}, they are exactly the graphs with no induced \(P_4\) or \(C_4\)~\cite{Golumbic1978}, and they form the smallest class containing \(K_1\) that is closed under disjoint union and under the addition of a universal vertex (\cref{def:triviallyPerfect-recursive})~\cite{YanChenChang}. They can be recognized in linear time~\cite{YanChenChang,Chu2008}, and this recognition produces the decomposition tree that our algorithm takes as input. Every threshold graph is trivially perfect, and every trivially perfect graph is a cograph.

Trivially perfect graphs are a good testing ground for the proper variant. Every trivially perfect graph is an interval graph~\cite{Brandstadt-book}, so its thinness is always~\(1\)~\cite{M-O-R-C-thinness}. Its proper thinness, instead, is unbounded: let \(\mathrm{claw}_1\) be the claw and, for \(k \ge 2\), let \(\mathrm{claw}_k\) be obtained from three disjoint copies of \(\mathrm{claw}_{k-1}\) by adding a universal vertex; these graphs are trivially perfect, and \(\pthin(\mathrm{claw}_k) = k+1\) (\cite[Proposition 10]{B-D-thinness}, restated here as \cref{obs:claw-pthin2}). In this class, then, the whole difficulty of the parameter comes from the extra requirement of the proper variant. Moreover, since computing the thinness of a cograph is polynomial~\cite{B-D-thinness} while the complexity of its proper thinness is open, these two classes are a natural first step towards the cograph case.

First, we determine the proper thinness of every threshold graph, and we show that it never exceeds~\(2\).

\begin{introtheorem}[proved as \cref{thm:threshold}]
Let \(G\) be a threshold graph. If \(G\) is claw-free, then \(\pthin(G) = 1\); otherwise, \(\pthin(G) = 2\).
\end{introtheorem}

For trivially perfect graphs the parameter can take larger values, and we characterize those of proper thinness at most~\(2\) in three complementary ways: structurally, algorithmically, and by forbidden induced subgraphs.

Disjoint unions are easy to handle, because the proper thinness of a disjoint union is the maximum of the proper thinness of its parts (\cite[Theorem 14]{B-D-thinness}, restated here as \cref{thm:union}). Universal extensions are not: the proper thinness of the graph \(\Gplusu\) obtained from a graph \(G\) by adding a universal vertex \(u\) is not determined by the proper thinness of the connected components of \(G\). For example, let \(H_1\) be the graph obtained by adding a universal vertex to the disjoint union of a claw and an isolated vertex (\cref{fig:stag-bnd}), and let \(H_2\) be the graph obtained by adding a universal vertex to two disjoint claws. Both have proper thinness~\(2\). If we add to each of them an isolated vertex and then a universal vertex, the graph obtained from \(H_1\) still has proper thinness~\(2\), whereas the one obtained from \(H_2\) is the \lotus{}, of proper thinness~\(3\) (\cref{lem:lotus_minimal}).

The difference between \(H_1\) and \(H_2\) lies in the room that their solutions leave in the ordering for the vertices that the next operation will add. To see which kind of room matters, consider a strongly consistent solution of \(\Gplusu\) with two classes, and let \(C_u\) be the class of \(u\). Every connected component of \(G\) of proper thinness~\(2\) has vertices in \(C_u\), and the vertices of \(C_u\) smaller than \(u\) form a clique, and so do those greater than \(u\) (\cref{prop:cliques}), so each of these two sets lies in a single component. With three or more components of proper thinness~\(2\) there is not enough room, and \(\pthin(\Gplusu)=3\). With exactly two, each of them needs a solution in which its universal vertex is the greatest vertex of its class, up to reversing the order; we call such a solution \emph{boundary} (\cref{def:boundary-solution}). With exactly one, the vertices of the other components have to be placed in the other class, between the vertices of \(C_u\) smaller than \(u\) and those greater than \(u\); this is possible precisely when the component has a solution that leaves a class free for them, which we call \emph{staggered} (\cref{def:staggered}). The graph \(H_1\) has an optimal staggered solution, and \(H_2\) does not. For a universal extension, we characterize when the resulting graph still has proper thinness at most~\(2\), and when it still admits an optimal staggered or an optimal boundary solution, in terms of the same properties of the connected components being combined, of their number, and of whether they are complete. This yields a recursive description of the parameter along the decomposition tree of the graph.

\begin{introtheorem}[proved as \cref{theorem:linearAlgorithm}]
There exists a linear-time algorithm that decides whether the proper thinness of a trivially perfect graph is at most~\(2\).
\end{introtheorem}

The algorithm computes \(\min(\pthin(G), 3)\) by a single traversal of the decomposition tree. Following the same recursion backwards from a minimal graph of proper thinness~\(3\), we obtain two obstructions: the \(\mathrm{claw}_2\) and the \lotus{} (\cref{fig:claw_2,fig:lotus}). The \(\mathrm{claw}_2\) comes from three components of proper thinness~\(2\), and the \lotus{} from a single component without an optimal staggered solution, while the remaining case, two components one of which has no optimal boundary solution, produces no new obstruction.

\begin{introtheorem}[proved as \cref{thm:forbidden-minimal}]
Let \(G\) be a trivially perfect graph. Then \(\pthin(G) \leq 2\) if and only if \(G\) contains neither the \(\mathrm{claw}_2\) nor the \lotus{} as an induced subgraph. Moreover, both graphs are minimal forbidden induced subgraphs for the class of trivially perfect graphs with proper thinness at most~\(2\).
\end{introtheorem}

After the preliminaries (\cref{sec:preliminaries}) and some general facts about strongly consistent solutions (\cref{sec:general-facts}), \cref{sec:threshold,sec:triviallyperfect,sec:forbidden} prove the three results above, respectively, and \cref{sec:conclusions} discusses open problems.

\section{Preliminaries}
\label{sec:preliminaries}

All graphs in this paper are finite, undirected, and have no loops or multiple edges.

\subsection{Basic notation and definitions}

Let \(G\) be a graph. We denote by \(V(G)\) its set of vertices and by \(E(G)\) its set of edges. The \emph{neighborhood} of a vertex \(v \in V(G)\), denoted by \(N(v)\), is the set of vertices adjacent to \(v\). The \emph{closed neighborhood} of \(v\) is \(N[v] = N(v) \cup \{v\}\). The \emph{degree} of \(v\), denoted by \(d(v)\), is the number of edges of \(G\) having \(v\) as an endpoint.

A graph \(H\) is a \emph{subgraph} of \(G\) if \(V(H) \subseteq V(G)\) and \(E(H) \subseteq E(G)\). Given a set \(W \subseteq V(G)\), the subgraph of \(G\) \emph{induced} by \(W\) is the graph \(G[W]\) with vertex set \(W\) and edge set consisting of all edges of \(G\) with both endpoints in \(W\). We write \(G - W\) for \(G[V(G) \setminus W]\). In particular, for a vertex $u \in V(G)$, we write $G-u$ for \(G[V(G) \setminus \{u\}]\).

Two graphs \(G\) and \(H\) are \emph{isomorphic} if there is a bijection between their vertex sets that preserves adjacencies; in that case we write \(G \cong H\). The \emph{disjoint union} of graphs \(G_1\) and \(G_2\), denoted \(G_1 \cup G_2\), is the graph whose vertex set is the disjoint union of \(V(G_1)\) and \(V(G_2)\), and whose edge set is \(E(G_1) \cup E(G_2)\).

A set of vertices \(W \subseteq V(G)\) is a \emph{clique} if every two distinct vertices in \(W\) are adjacent. A graph is \emph{complete} if its vertex set is a clique; we denote by \(K_n\) the complete graph on \(n\) vertices. A vertex \(v \in V(G)\) is \emph{universal} if it is adjacent to every other vertex of \(G\), i.e., \(N[v] = V(G)\).

A graph \(H\) is a \emph{forbidden induced subgraph} for a class \(\mathcal{C}\) of graphs if no graph in \(\mathcal{C}\) contains \(H\) as an induced subgraph. The forbidden subgraph is \emph{minimal} if every proper induced subgraph of \(H\) belongs to \(\mathcal{C}\).

\subsection{Interval graphs and proper interval graphs}

An \emph{interval graph} is the intersection graph of a family of intervals on the real line~\cite{Brandstadt-book}, and a \emph{proper interval graph} is an interval graph that admits such a model in which no interval is properly contained in another~\cite{Rob-uig}. Interval graphs are exactly the graphs of thinness \(1\)~\cite{M-O-R-C-thinness}, and proper interval graphs are exactly the graphs of proper thinness \(1\)~\cite{B-D-thinness}, since their characterizations by vertex orderings~\cite{Ola-interval,R-PR-interval} and~\cite{L-O-prop-int,D-H-H-circ-arc}, respectively, coincide with \cref{def:thinness,def:proper-thinness} below for \(k=1\). We also use the following characterization of proper interval graphs, where the \emph{claw} \(K_{1,3}\) is the graph obtained by adding a universal vertex to three isolated vertices.

\begin{proposition}[{\cite{Rob-uig}}]\label{th:Roberts}
A graph is a proper interval graph if and only if it is an interval graph and does not contain the claw \(K_{1,3}\) as an induced subgraph.
\end{proposition}

\subsection{Thinness and proper thinness}
We now recall the definitions of thinness and proper thinness. Given a graph \(G\), a \emph{solution} of \(G\) is a pair \(\Gamma = (S,<)\), where \(S\) is a partition of \(V(G)\) and \(<\) is an ordering of \(V(G)\).

\begin{definition}[{\cite{M-O-R-C-thinness}}]\label{def:thinness}
A graph \(G = (V, E)\) is \emph{\(k\)-thin} if there exists an ordering \(v_1 < v_2 < \dots < v_n\) of \(V\) and a partition of \(V\) into \(k\) classes such that, for every triple \((r, s, t)\) with \(r < s < t\), if \(v_r\) and \(v_s\) belong to the same class and \(v_r v_t \in E\), then \(v_s v_t \in E\).

An ordering and a partition satisfying this property are said to be \emph{consistent}, and so is the corresponding solution. The minimum \(k\) for which \(G\) is \(k\)-thin is the \emph{thinness} of \(G\), denoted by \(\thin(G)\).
\end{definition}

\begin{definition}[{\cite{B-D-thinness}}]\label{def:proper-thinness}
A graph \(G = (V, E)\) is \emph{proper \(k\)-thin} if there exist an ordering \(v_1 < v_2 < \dots < v_n\) of \(V\) and a partition of \(V\) into \(k\) classes such that both the ordering \(v_1, v_2, \dots, v_n\) and its reverse \(v_n, \dots, v_1\) are consistent with the partition (see \cref{def:thinness}). Equivalently, for every triple \((r, s, t)\) with \(r < s < t\):
\begin{itemize}
    \item[\hypertarget{sc:cond-i}{(i)}] if \(v_r\) and \(v_s\) belong to the same class and \(v_r v_t \in E\), then \(v_s v_t \in E\); and
    \item[\hypertarget{sc:cond-ii}{(ii)}] if \(v_s\) and \(v_t\) belong to the same class and \(v_r v_t \in E\), then \(v_r v_s \in E\).
\end{itemize}
In this case, the corresponding solution \(\Gamma = (S, <)\) is said to be \emph{strongly consistent}. The minimum \(k\) for which \(G\) admits such a representation is the \emph{proper thinness} of \(G\), denoted by \(\pthin(G)\).
\end{definition}

Examples of a consistent solution and of a strongly consistent solution are shown in \cref{fig:thinness,fig:properthinness}, respectively. A strongly consistent solution of \(G\) is \emph{optimal} if it has exactly \(\pthin(G)\) classes.

\medskip
\noindent\textbf{Drawing conventions.} In our figures, the vertices of a
solution \(\Gamma = (S,<)\) are placed horizontally in increasing order of \(<\): if \(u < v\), then the horizontal coordinate of \(u\) is smaller than that of \(v\). Each class of \(S\) is drawn as a
separate row enclosed in a dashed box labeled with the name of the class (see, e.g., \cref{fig:thinness,fig:properthinness}). When the
graph is disconnected and its connected components are relevant, the vertices are
colored according to their component.

\medskip
We use the following three known facts about proper thinness, where \(\Gplusu\) denotes the graph obtained from a graph \(G\) by adding a new
universal vertex \(u\).

\begin{theorem}[{{\cite[Theorem 14]{B-D-thinness}}}]
\label{thm:union}
  Let $G_1$ and $G_2$ be graphs. Then \[\pthin(G_1\cup G_2)=
  \max\{\pthin(G_1),\pthin(G_2)\}.\]
\end{theorem}

The following theorem, stated by \citet{B-D-thinness} for joins of graphs, is specialized here to the addition of a single universal vertex.
\begin{theorem}[{{\cite[Theorem 15]{B-D-thinness}}}]
\label{thm:join}
  Let $G$ be a graph. Then \[\pthin(\Gplusu)\le\pthin(G)+1.\]
\end{theorem}

\begin{theorem}[{\cite{B-D-thinness,thinness-graph-families}}]\label{thm:hereditary}
  If \(H\) is an induced subgraph of a graph \(G\), then \[\pthin(H) \le \pthin(G).\]
\end{theorem}

\subsection{Threshold graphs}

Threshold graphs are built by adding isolated and universal vertices, as follows.

\begin{definition}
\label{def:threshold}
    A \emph{threshold graph} is a graph that can be constructed from a single vertex by repeatedly applying the following two operations:
    \begin{itemize}
        \item adding a new isolated vertex; and
        \item adding a new universal vertex.
    \end{itemize}
\end{definition}
Equivalently, threshold graphs are exactly the graphs that contain no induced \(P_4\), \(C_4\), or \(2K_2\)~\cite{Ch-H-thres,Ch-H-aggregation}. The sequence in which the vertices are added is a \emph{threshold order} of the graph; it plays a central role in \cref{sec:threshold}.

\subsection{Trivially perfect graphs}

We use the following recursive characterization of trivially perfect graphs.

\begin{definition}\label{def:triviallyPerfect-recursive}
The class of \emph{trivially perfect graphs} is defined recursively as follows:
\begin{itemize}
    \item a single vertex is trivially perfect;
    \item the disjoint union of two trivially perfect graphs is trivially perfect;
    \item if \(G\) is trivially perfect, then the graph obtained by adding a new universal vertex to \(G\) is trivially perfect.
\end{itemize}
\end{definition}

Equivalently, a graph is trivially perfect if and only if every connected induced subgraph of it has a universal vertex~\cite{Brandstadt-book}; in particular, every connected trivially perfect graph has a universal vertex. Moreover, trivially perfect graphs are exactly the graphs that contain no induced \(P_4\) or \(C_4\)~\cite{Golumbic1978}. An example of a trivially perfect graph, together with the tree of operations that builds it, is shown in \cref{fig:tree}. Every trivially perfect graph is an interval graph~\cite{Brandstadt-book}, and an interval graph is a proper interval graph if and only if it is claw-free (\cref{th:Roberts}). Since the graphs of proper thinness \(1\) are exactly the proper interval graphs, we obtain the following.

\begin{lemma}[Trivially perfect graphs of proper thinness 1]
\label{lem:claw-free-pthin1}
A trivially perfect graph has proper thinness \(1\) if and only if it is claw-free. Equivalently, it admits a proper interval representation. Moreover, this property can be tested in linear time~\cite{F-M-M-prop-int}.
\end{lemma}

\section{General facts about strongly consistent solutions}
\label{sec:general-facts}

The results of this section hold for arbitrary graphs and are used repeatedly in the rest of the paper. They describe how a strongly consistent solution behaves when passing to an induced subgraph, reversing the order, or adding a universal vertex, and how strong consistency can be checked component by component.

\begin{definition}
\label{def:restriction}
    Let $G$ be a graph, let $\Gamma = (S,<)$ be a solution of $G$, and let $H$ be an induced subgraph of $G$.
    The \emph{restriction} of \(\Gamma\) to \(V(H)\) is the solution \((S', <')\) of \(H\), where \(S' = \{\, C \cap V(H) : C \in S,\; C \cap V(H) \neq \emptyset \,\}\) and \(<'\) is the restriction of \(<\) to \(V(H)\); that is, it is obtained from \(\Gamma\) by deleting the vertices outside \(V(H)\) and keeping, for the remaining vertices, their classes and their relative order.
\end{definition}

Strong consistency is inherited by restrictions.

\begin{observation}
\label{obs:restriction-consistent}
Let \(G\) be a graph, let \(\Gamma\) be a strongly consistent solution of \(G\), and let \(H\) be an induced subgraph of \(G\). Then the restriction of \(\Gamma\) to \(V(H)\) is a strongly consistent solution of \(H\).
\end{observation}

\begin{proof}
Every triple of vertices of \(H\) is also a triple of vertices of \(G\), with the same relative order and the same class membership, and two vertices of \(H\) are adjacent in \(H\) if and only if they are adjacent in \(G\). Hence both conditions of \cref{def:proper-thinness} are inherited.
\end{proof}

Since reversing the order exchanges the two conditions of \cref{def:proper-thinness}, we have the following.

\begin{observation}
\label{prop:symmetry}
Let \(G\) be a graph, let \(S\) be a partition of \(V(G)\), and let \(<\) be an ordering of \(V(G)\). Then \((S,<)\) is a strongly consistent solution of \(G\) if and only if \((S,> )\) is a strongly consistent solution of \(G\), where \(>\) denotes the reverse of \(<\).
\end{observation}

Most of the constructions in this paper insert a universal vertex \(u\) into a strongly consistent solution that has already been built. The next lemma isolates what has to be checked when doing so. Among the conditions of \cref{def:proper-thinness} on the triples containing \(u\), only two have a conclusion that is not an edge incident to \(u\) (the others hold trivially, since \(u\) is universal): condition~\condi{} when \(u\) is the smallest vertex of the triple, and condition~\condii{} when it is the greatest. In both, the hypothesis forces the middle vertex to lie in the class of \(u\), which is why \cref{cond:ins-above,cond:ins-below} below quantify only over \(C_u\).

\begin{lemma}[Insertion of a universal vertex]
\label{lem:universal-insertion}
Let \(H\) be a graph, let \(u\) be a universal vertex of \(H\), let \(\Gamma = (S,<)\) be a solution of \(H\), and let \(C_u \in S\) be the class containing \(u\). Then \(\Gamma\) is strongly consistent if and only if the following three conditions hold:
\begin{enumerate}
    \item\label[condition]{cond:ins-restriction} the restriction of \(\Gamma\) to \(V(H) \setminus \{u\}\) (\cref{def:restriction}) is a strongly consistent solution of \(H - u\);
    \item\label[condition]{cond:ins-above} every \(v_s \in C_u\) with \(u < v_s\) is adjacent to every vertex \(v_t\) of \(H\) with \(v_s < v_t\);
    \item\label[condition]{cond:ins-below} every \(v_s \in C_u\) with \(v_s < u\) is adjacent to every vertex \(v_r\) of \(H\) with \(v_r < v_s\).
\end{enumerate}
\end{lemma}

\begin{proof}
\noindent\textbf{(\(\Rightarrow\))} \Cref{cond:ins-restriction} is \cref{obs:restriction-consistent}. For \cref{cond:ins-above}, let \(v_s \in C_u\) with \(u < v_s\) and let \(v_t\) be a vertex of \(H\) with \(v_s < v_t\). The triple \((u, v_s, v_t)\) satisfies \(u < v_s < v_t\), the vertices \(u\) and \(v_s\) belong to the same class \(C_u\), and \(u v_t \in E(H)\) because \(u\) is universal; condition~\condi{} applied to this triple yields \(v_s v_t \in E(H)\). For \cref{cond:ins-below}, let \(v_s \in C_u\) with \(v_s < u\) and let \(v_r\) be a vertex of \(H\) with \(v_r < v_s\). The triple \((v_r, v_s, u)\) satisfies \(v_r < v_s < u\), the vertices \(v_s\) and \(u\) belong to the same class \(C_u\), and \(v_r u \in E(H)\) because \(u\) is universal; condition~\condii{} applied to this triple yields \(v_r v_s \in E(H)\).

\medskip

\noindent\textbf{(\(\Leftarrow\))} Let \(v_r < v_s < v_t\) be a triple of vertices of \(H\). If it does not contain \(u\), both conditions of \cref{def:proper-thinness} follow from \cref{cond:ins-restriction}, since \(H-u\) is an induced subgraph of \(H\). If \(v_s = u\), both conclusions \(v_s v_t\) and \(v_r v_s\) are edges because \(u\) is universal. If \(v_r = u\), the conclusion \(v_r v_s\) of~\condii{} is an edge for the same reason, and if the hypothesis of~\condi{} holds, then \(v_s \in C_u\) and \(v_s > u\), so \cref{cond:ins-above} gives its conclusion \(v_s v_t \in E(H)\). The case \(v_t = u\) is symmetric, exchanging the roles of~\condi{} and~\condii{} and using \cref{cond:ins-below}.
\end{proof}

Note that when \(u\) is the greatest vertex of its class,
\cref{cond:ins-above} is vacuous, and when \(C_u = \{u\}\), both conditions are vacuous and the
lemma reduces to \cref{cond:ins-restriction}. 

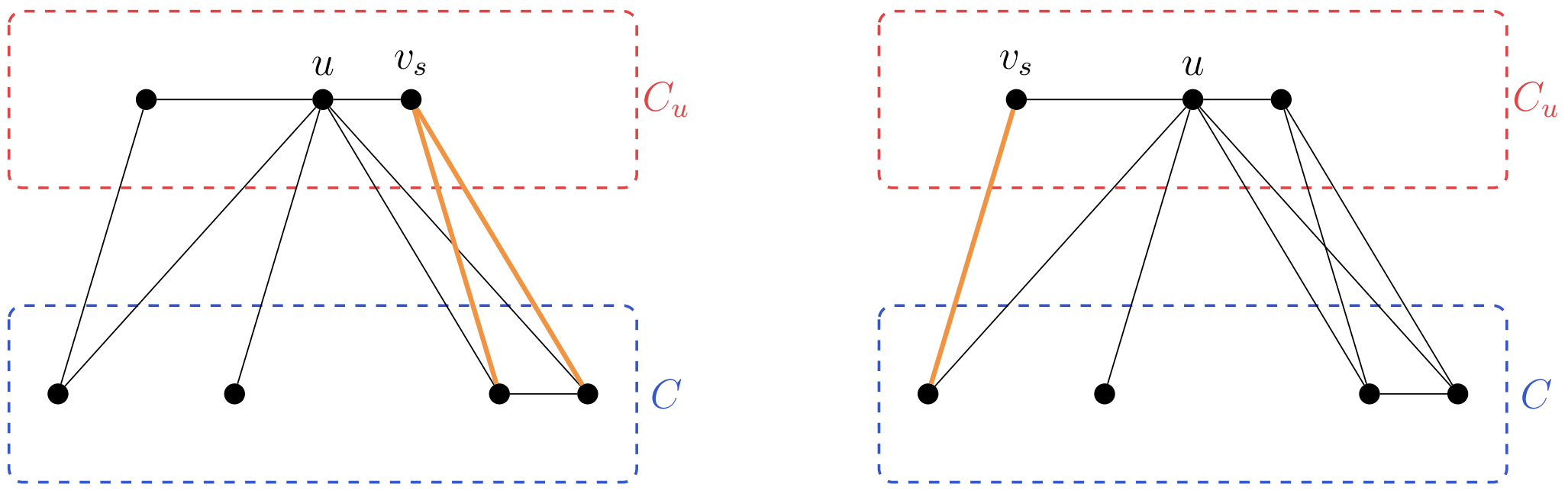
\begin{figure}[H]
\centering
\begin{subfigure}[b]{0.45\textwidth}
\centering
\scalebox{0.85}{
\begin{tikzpicture}[x={(0cm,-1cm)}, y={(1cm,0cm)}, blackv/.style={draw, circle, fill=black, inner sep=2pt},
    highlighted/.style={palResaltado, line width=1.4pt}]
    \node[blackv] (y1) at (0,0.9) {};
    \node[blackv, label=above:$u$] (u) at (0,2.7) {};
    \node[blackv, label=above:$v_s$] (y2) at (0,3.6) {};
    \draw[dashed, rounded corners, palClassA, thick] (-0.9,-0.5) rectangle (0.9,5.9);
    \node[palClassA] at (0,6.2) {$C_u$};
    \node[blackv] (x1) at (3,0) {};
    \node[blackv] (x2) at (3,1.8) {};
    \node[blackv] (x3) at (3,4.5) {};
    \node[blackv] (x4) at (3,5.4) {};
    \draw[dashed, rounded corners, palClassB, thick] (2.1,-0.5) rectangle (3.9,5.9);
    \node[palClassB] at (3,6.2) {$C$};
    \draw (u) to (y1);
    \draw (u) to (y2);
    \draw (u) to (x1);
    \draw (u) to (x2);
    \draw (u) to (x3);
    \draw (u) to (x4);
    \draw (y1) to (x1);
    \draw (x3) to (x4);
    \draw[highlighted] (y2) to (x3);
    \draw[highlighted] (y2) to (x4);
\end{tikzpicture}
}
\caption{\Cref{cond:ins-above}.}
\label{fig:insertion-a}
\end{subfigure}
\hfill
\begin{subfigure}[b]{0.45\textwidth}
\centering
\scalebox{0.85}{
\begin{tikzpicture}[x={(0cm,-1cm)}, y={(1cm,0cm)}, blackv/.style={draw, circle, fill=black, inner sep=2pt},
    highlighted/.style={palResaltado, line width=1.4pt}]
    \node[blackv, label=above:$v_s$] (y1) at (0,0.9) {};
    \node[blackv, label=above:$u$] (u) at (0,2.7) {};
    \node[blackv] (y2) at (0,3.6) {};
    \draw[dashed, rounded corners, palClassA, thick] (-0.9,-0.5) rectangle (0.9,5.9);
    \node[palClassA] at (0,6.2) {$C_u$};
    \node[blackv] (x1) at (3,0) {};
    \node[blackv] (x2) at (3,1.8) {};
    \node[blackv] (x3) at (3,4.5) {};
    \node[blackv] (x4) at (3,5.4) {};
    \draw[dashed, rounded corners, palClassB, thick] (2.1,-0.5) rectangle (3.9,5.9);
    \node[palClassB] at (3,6.2) {$C$};
    \draw (u) to (y1);
    \draw (u) to (y2);
    \draw (u) to (x1);
    \draw (u) to (x2);
    \draw (u) to (x3);
    \draw (u) to (x4);
    \draw (y2) to (x3);
    \draw (y2) to (x4);
    \draw (x3) to (x4);
    \draw[highlighted] (y1) to (x1);
\end{tikzpicture}
}
\caption{\Cref{cond:ins-below}.}
\label{fig:insertion-b}
\end{subfigure}
\caption{The two conditions of \cref{lem:universal-insertion} on a solution with two classes, where \(u\) is a universal vertex: a vertex \(v_s \in C_u\) with \(v_s > u\) must be adjacent to every vertex \(v_t > v_s\), and a vertex \(v_s \in C_u\) with \(v_s < u\) must be adjacent to every vertex \(v_r < v_s\). The required edges are highlighted.}
\label{fig:insertion}
\end{figure}

\Cref{fig:insertion} illustrates \cref{cond:ins-above,cond:ins-below}.

The next lemma applies \cref{lem:universal-insertion} to the simplest way of adding
a universal vertex: placing it next to another universal vertex, in its class.

\begin{lemma}
\label{lem:twin-insertion}
Let \(G\) be a graph, let \(x\) be a universal vertex of \(G\), and let
\(\Gamma = (S,<)\) be a strongly consistent solution of \(G\). Let \(\Gamma'\) be
obtained from \(\Gamma\) by placing the new vertex \(u\) of \(\Gplusu\) in the class \(C\)
of \(x\), greater than \(x\) and smaller than every vertex of \(C\) that is greater
than \(x\), and keeping the relative order of all the other vertices. Then
\(\Gamma'\) is a strongly consistent solution of \(\Gplusu\) with the same number of
classes as \(\Gamma\). The same holds if \(u\) is placed smaller than \(x\) and greater
than every vertex of \(C\) that is smaller than \(x\).
\end{lemma}

\begin{proof}
We apply \cref{lem:universal-insertion} to the universal vertex \(u\) of \(\Gplusu\),
whose class in \(\Gamma'\) is \(C\). \Cref{cond:ins-restriction} holds because the restriction
of \(\Gamma'\) to \(V(G)\) is exactly \(\Gamma\). By construction, we have that \(x < u\), and no vertex of \(C\)
lies between \(x\) and \(u\). Hence the vertices of \(C\) greater than \(u\) are those greater than \(x\), and the vertices of \(C\) smaller than \(u\) are \(x\), which is universal, and those smaller than \(x\). Therefore, applying the same \cref{lem:universal-insertion} to the universal vertex $x$ yields the other two conditions of \cref{lem:universal-insertion} for $u$. For the second statement, reversing the order of \(\Gamma\) and of \(\Gamma'\) (\cref{prop:symmetry}) turns this placement of \(u\) into the one of the first statement. Finally, \(u\) is added to an existing class, so the number of classes does not change.
\end{proof}

Another consequence of \cref{lem:universal-insertion} is that, in a strongly consistent solution, the vertices of the class of a universal vertex that are smaller than it form a clique, and so do those greater than it.

\begin{lemma}
\label{prop:cliques}
Let $G$ be a graph, let \( u \) be a universal vertex of \( G \) and let \( C \) be the class of a strongly consistent solution of $G$ that contains \(u\). All vertices \( s < u \) of \( C \) form a clique, and all vertices \( t > u \) of \( C \) form a clique.
\end{lemma}

\begin{proof}
By \cref{cond:ins-above} of \cref{lem:universal-insertion}, every vertex of \(C\) greater than \(u\) is adjacent to every vertex of \(G\) greater than it; in particular, it is adjacent to every vertex of \(C\) greater than it, so the vertices of \(C\) greater than \(u\) are pairwise adjacent. The statement for the vertices of \(C\) smaller than \(u\) follows from \cref{cond:ins-below} in the same way.
\end{proof}

Finally, when a solution is strongly consistent on each connected component, only the triples that cross components can violate strong consistency, and there are none if every component is consecutive in the order.

\begin{lemma}
\label{lem:components}
Let \(G\) be a graph with connected components \(\CompOne, \dots, Comp_m\), and let \(\Gamma = (S,<)\) be a solution of \(G\) such that the restriction of \(\Gamma\) to \(V(Comp_i)\) is strongly consistent for every \(i\). If a triple \(v_r < v_s < v_t\) violates condition~\condi{} or~\condii{} of \cref{def:proper-thinness}, then \(v_r\) and \(v_t\) belong to the same connected component and \(v_s\) belongs to a different one.
\end{lemma}

\begin{proof}
The hypothesis of both conditions includes \(v_r v_t \in E(G)\), so a violating triple has \(v_r\) and \(v_t\) in the same connected component, say \(Comp_j\). If \(v_s\) also belonged to \(Comp_j\), the whole triple would lie in \(Comp_j\), and the restriction of \(\Gamma\) to \(V(Comp_j)\) is strongly consistent, a contradiction.
\end{proof}

\begin{lemma}
\label{cor:contiguous-components}
Under the hypotheses of \cref{lem:components}, if the vertices of each connected component are consecutive in \(<\), then \(\Gamma\) is a strongly consistent solution of \(G\).
\end{lemma}

\begin{proof}
If \(v_r\) and \(v_t\) belong to the same connected component and \(v_r < v_s < v_t\), then \(v_s\) belongs to that component as well, because its vertices are consecutive in \(<\). Hence no triple can be as in \cref{lem:components}.
\end{proof}

We end this section by transferring a solution of \(\Gplusu\) to a connected component of \(G\), letting a universal vertex of it take the place of \(u\).

\begin{lemma}[Transfer]
\label{lem:transfer}
Let \(\Gamma=(S,<)\) be a strongly consistent solution of \(\Gplusu\), and let \(u_1\) be a universal vertex of a connected component \(Comp\) of \(G\). Let \(\Gamma_1\) be the solution of \(Comp\) obtained from the restriction of \(\Gamma\) to \((V(Comp) \setminus \{u_1\}) \cup \{u\}\) by renaming \(u\) as \(u_1\). Then \(\Gamma_1\) is strongly consistent with at most \(|S|\) classes, and, letting \(u\) correspond to \(u_1\) and every other vertex to itself, two vertices of \(Comp\) are in the same class of \(\Gamma_1\) if and only if the corresponding vertices are in the same class of \(\Gamma\), and one is smaller than the other in \(\Gamma_1\) if and only if the same holds for the corresponding vertices in \(\Gamma\). In particular, the class of \(u_1\) in \(\Gamma_1\) consists of \(u_1\) and the vertices of \((C_u \cap V(Comp)) \setminus \{u_1\}\), where \(C_u\) is the class of \(u\) in \(S\), and every other class of \(\Gamma_1\) is contained in a class of \(S\) different from \(C_u\).
\end{lemma}

\begin{proof}
Since \(u\) and \(u_1\) have the same neighbors in \(Comp - u_1\), the renaming is an isomorphism from the subgraph of \(\Gplusu\) induced by \((V(Comp) \setminus \{u_1\}) \cup \{u\}\) onto \(Comp\). Apply \cref{obs:restriction-consistent}.
\end{proof}

\section{Proper thinness of threshold graphs}
\label{sec:threshold}

In this section we characterize the proper thinness of threshold graphs (\cref{def:threshold}). Since threshold graphs are the graphs with no induced \(P_4\), \(C_4\), or \(2K_2\)~\cite{Ch-H-thres,Ch-H-aggregation}, and trivially perfect graphs are the graphs with no induced \(P_4\) or \(C_4\)~\cite{Golumbic1978}, every threshold graph is trivially perfect.

For the graphs $\mathrm{claw}_k$ defined in the Introduction, \citet{B-D-thinness} prove the following.

\begin{observation}[{{\cite[Proposition 10]{B-D-thinness}}}]
\label{obs:claw-pthin2}
For every \(k \geq 1\), \(\pthin(\mathrm{claw}_k) = k + 1\). In particular, the claw
has proper thinness \(2\) and \(\mathrm{claw}_2\) has proper thinness \(3\).
\end{observation}

We now characterize the proper thinness of threshold graphs.

\begin{theorem}
\label{thm:threshold}
 Let \( G \) be a threshold graph. If \( G \) is claw-free, then \( G \) has proper thinness equal to 1. On the other hand, if \( G \) contains an induced claw, then \( G \) has proper thinness equal to 2.
\end{theorem}

\begin{proof}
If $G$ is claw-free, then $G$ has proper thinness $1$ by \cref{lem:claw-free-pthin1}, since every threshold graph is trivially perfect.

Suppose then that $G$ contains an induced claw. Since the claw has proper thinness 2 by \cref{obs:claw-pthin2}, $G$ has proper thinness at least 2 by \cref{thm:hereditary}. To show that the proper thinness is at most 2, we build a solution $\Gamma = (S, <)$ where $<$ is a threshold order and $S$ has two classes $C_1$ and $C_2$: place the vertices added as isolated vertices in class $C_1$ and the vertices that were added as universal vertices in class $C_2$. The first vertex of the threshold order is regarded as added as an isolated vertex, so it belongs to $C_1$.

\noindent\textbf{Strong consistency of $\Gamma$.} Consider three vertices $v_r, v_s, v_t$ where $v_r < v_s < v_t$; we verify conditions~\condi{} and~\condii{} of \cref{def:proper-thinness} according to the class of $v_t$. If $v_t \in C_1$, then $v_t$ was added as an isolated vertex and $v_r < v_t$ in the threshold order, so $v_r v_t \notin E(G)$ and neither condition applies. If $v_t \in C_2$, then $v_t$ was added as a universal vertex and $v_r, v_s < v_t$ in the threshold order, so $v_t$ is adjacent to both of them; in particular $v_s v_t \in E(G)$, so condition~\condi{} holds. For condition~\condii{}, if $v_s$ and $v_t$ belong to the same class, then $v_s \in C_2$, so $v_s$ was added as a universal vertex and $v_r < v_s$, hence $v_r v_s \in E(G)$. Therefore $G$ has proper thinness exactly $2$.
\end{proof}

Since threshold graphs are trivially perfect and, by \cref{thm:threshold}, have proper thinness at most 2, the linear-time algorithm of \cref{sec:triviallyperfect}, which computes \(\min(\pthin(G),3)\), computes the proper thinness of a threshold graph in linear time.

\section{Proper thinness of trivially perfect graphs}
\label{sec:triviallyperfect}

In this section we characterize the trivially perfect graphs of proper thinness at most 2 and give a linear-time algorithm that recognizes them. The algorithm processes the tree of operations bottom-up, keeping for each connected subgraph its proper thinness and whether it admits an optimal \emph{staggered} (\cref{def:staggered}) or \emph{boundary} (\cref{def:boundary-solution}) solution.

\subsection{Characterization of proper thinness at most 2}

In this subsection, we characterize the trivially perfect graphs with proper thinness at most 2. We first introduce some notation and auxiliary definitions, also used by the linear-time algorithm of the next subsection.

Given a vertex \(x\) and a set of vertices \(A\), we write \(x < A\) to mean that \(x < a\) for every \(a \in A\), and \(x > A\) to mean that \(x > a\) for every \(a \in A\). Similarly, for two sets of vertices \(A\) and \(B\), we write \(A < B\) to mean that \(a < b\) for every \(a \in A\) and every \(b \in B\).

Recall that \(\Gplusu\) denotes the graph obtained from \(G\) by adding a new universal vertex \(u\); we call this operation the \emph{universal extension} of \(G\).

When a universal vertex is added to a trivially perfect graph, the proper thinness of the resulting graph is not determined solely by the proper thinness of its connected components: it also depends on whether their optimal solutions leave room in the order for the new universal vertex and for the vertices of the other components. The next two definitions describe the two kinds of solutions that leave such room, and the characterization below relies on whether each connected component admits optimal solutions of these kinds.

\begin{definition}[Staggered solution]
\label{def:staggered}
Let $G$ be a trivially perfect graph and let $\Gamma = (S,<)$ be a strongly consistent solution of $G$.
We say that $\Gamma$ is \emph{staggered} if there exists a universal vertex $u$ of $G$ belonging to a class $C \in S$ for which the staggered condition holds, i.e., letting
\[
A = \{ x \in C : x < u \},
\]
there exists a class $C_1 \in S$, different from $C$, such that for every vertex $w \in C_1$,
\begin{itemize}
    \item \hypertarget{prop1_staggered}{(i)} $w > A$, and
    \item \hypertarget{prop2_staggered}{(ii)} $w$ is not adjacent to any vertex of $A$.
\end{itemize}
In this case, we also say that $\Gamma$ is \emph{staggered with respect to} $u$.

\end{definition}

\begin{definition}[Boundary solution]
\label{def:boundary-solution}
Let \( G \) be a trivially perfect graph, and let \( \Gamma = (S, <) \) be a strongly consistent solution of \( G \). Let \( u \) be a universal vertex of \( G \) that belongs to the class \( C \) of \( S \).
The strongly consistent solution \( \Gamma \) is called a \emph{boundary} solution with respect to $u$ if for all \( w \in C \) with \( w \neq u \), we have \( u > w \) in the order \( < \).
\end{definition}

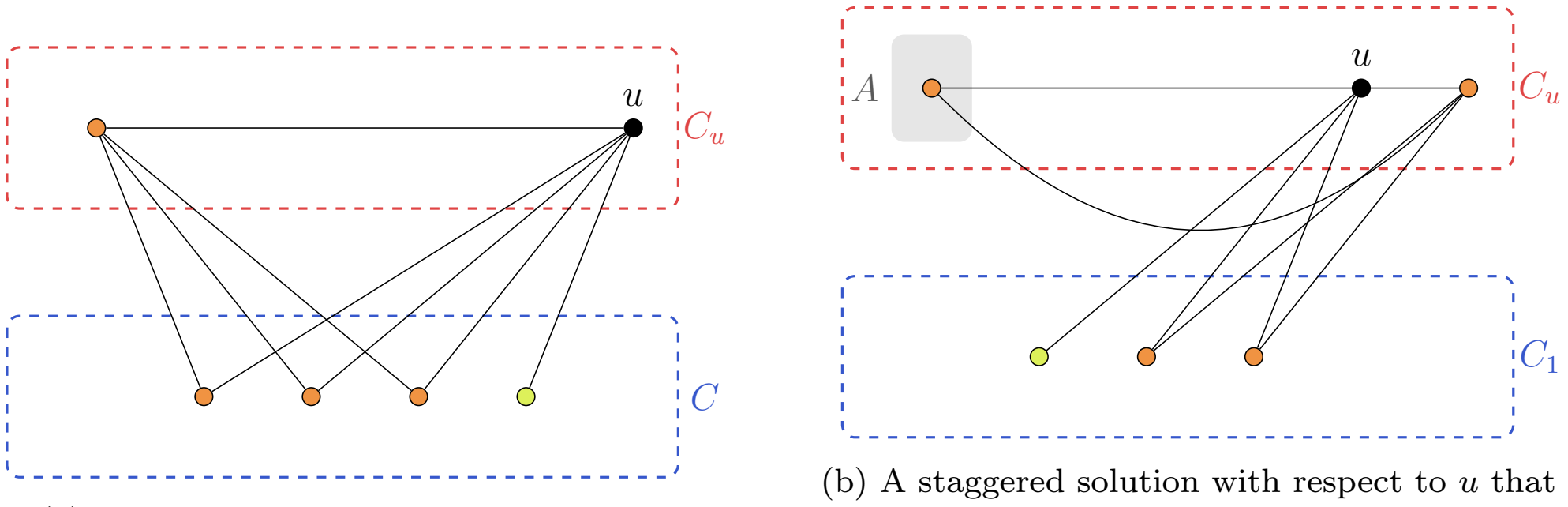
\begin{figure}[ht]
\centering
\begin{subfigure}[b]{0.49\textwidth}
\centering
\scalebox{0.75}{
\begin{tikzpicture}[x={(0cm,-1cm)}, y={(1cm,0cm)}, redv/.style={draw, circle, fill=palCompA, inner sep=2pt},
    bluev/.style={draw, circle, fill=palCompE, inner sep=2pt},
    blackv/.style={draw, circle, fill=black, inner sep=2pt}]
    \node[redv] (c) at (0,0) {};
    \node[blackv, label=above:$u$] (u) at (0,6) {};
    \draw[dashed, rounded corners, palClassA, thick] (-0.9,-1) rectangle (0.9,6.5);
    \node[palClassA] at (0,6.8) {$C_u$};

    \node[redv] (l1) at (3,1.2) {};
    \node[redv] (l2) at (3,2.4) {};
    \node[redv] (l3) at (3,3.6) {};
    \node[bluev] (z) at (3,4.8) {};
    \draw[dashed, rounded corners, palClassB, thick] (2.1,-1) rectangle (3.9,6.5);
    \node[palClassB] at (3,6.8) {$C$};

    \draw (u) to (c);
    \draw (u) to (l1);
    \draw (u) to (l2);
    \draw (u) to (l3);
    \draw (u) to (z);
    \draw (c) to (l1);
    \draw (c) to (l2);
    \draw (c) to (l3);
\end{tikzpicture}
}
\caption{A boundary solution with respect to \(u\).}
\label{fig:stag-bnd-a}
\end{subfigure}
\hfill
\begin{subfigure}[b]{0.49\textwidth}
\centering
\scalebox{0.75}{
\begin{tikzpicture}[x={(0cm,-1cm)}, y={(1cm,0cm)}, redv/.style={draw, circle, fill=palCompA, inner sep=2pt},
    bluev/.style={draw, circle, fill=palCompE, inner sep=2pt},
    blackv/.style={draw, circle, fill=black, inner sep=2pt}]
    \begin{pgfonlayer}{background}
        \fill[gray!20, rounded corners] (-0.6,-0.45) rectangle (0.6,0.45);
    \end{pgfonlayer}
    \node[gray!70!black] at (0,-0.75) {$A$};

    \node[redv] (l1) at (0,0) {};
    \node[blackv, label=above:$u$] (u) at (0,4.8) {};
    \node[redv] (c) at (0,6) {};
    \draw[dashed, rounded corners, palClassA, thick] (-0.9,-1) rectangle (0.9,6.5);
    \node[palClassA] at (0,6.8) {$C_u$};

    \node[bluev] (z) at (3,1.2) {};
    \node[redv] (l2) at (3,2.4) {};
    \node[redv] (l3) at (3,3.6) {};
    \draw[dashed, rounded corners, palClassB, thick] (2.1,-1) rectangle (3.9,6.5);
    \node[palClassB] at (3,6.8) {$C_1$};

    \draw (u) to (l1);
    \draw (u) to (z);
    \draw (u) to (l2);
    \draw (u) to (l3);
    \draw (u) to (c);
    \draw [bend left=45, looseness=1.25] (c) to (l1);
    \draw (c) to (l2);
    \draw (c) to (l3);
\end{tikzpicture}
}
\caption{A staggered solution with respect to \(u\) that is not boundary.}
\label{fig:stag-bnd-b}
\end{subfigure}
\caption{Two optimal strongly consistent solutions of the graph obtained by adding a universal vertex \(u\) to the disjoint union of a claw and an isolated vertex. In (b), the set \(A\) is shaded, and the class \(C_1\) witnesses that the solution is staggered. The solution in (a) is not staggered with respect to \(u\), but its reverse is (\cref{boundary_implies_staggered}).}
\label{fig:stag-bnd}
\end{figure}

\Cref{fig:stag-bnd} illustrates both definitions.

The next lemma, a direct consequence of \cref{lem:twin-insertion}, states that adding a universal vertex to a connected trivially perfect graph does not change its proper thinness; it is used repeatedly when constructing solutions for \(\Gplusu\).

\begin{lemma}[Universal extension of a connected graph]
\label{prop:many_universal_vertices}
For every connected \newline trivially perfect graph $G$, we have
\[\pthin(\Gplusu)=\pthin(G).\]
\end{lemma}
\begin{proof}
Every connected trivially perfect graph has a universal vertex~\cite{Brandstadt-book}; let $x$ be one. Applying \cref{lem:twin-insertion} to an optimal strongly consistent solution of $G$ and its universal vertex $x$, we obtain a strongly consistent solution of $\Gplusu$ with $\pthin(G)$ classes, so $\pthin(\Gplusu)\leq\pthin(G)$. Since $G$ is an induced subgraph of $\Gplusu$, we also have $\pthin(G)\leq\pthin(\Gplusu)$ by \cref{thm:hereditary}.
\end{proof}

Adding a universal vertex to a trivially perfect graph that already has one does not affect the existence of boundary or staggered solutions.

\begin{lemma}
    \label{lem:boundary-and-staggered-preserved}
    Let $G$ be a connected trivially perfect graph. Then $\Gplusu$ has a boundary or staggered solution on $p$ classes if and only if $G$ has a boundary or staggered solution on $p$ classes, respectively.
\end{lemma}
\begin{proof}
    Note that every connected trivially perfect graph $G$ has a universal vertex $u'$.

    Given a boundary solution for $G$ with respect to $u'$, we build one for $\Gplusu$ with respect to $u$ with the same number of classes by applying \cref{lem:twin-insertion}, adding $u$ right after $u'$ in its same class. To obtain a staggered solution for $\Gplusu$ from a staggered solution for $G$, we instead insert $u$ right before $u'$ in the same class. It can be verified that these solutions meet the definitions of staggered and boundary solutions (\cref{def:staggered,def:boundary-solution}), respectively.

    On the other hand, given a boundary or staggered solution $\Gamma$ for $\Gplusu$, we can obtain a boundary or staggered solution for $G$, respectively, by removing vertex $u'$ from $\Gamma$, and renaming $u$ to $u'$.
\end{proof}

The following lemma states that, for trivially perfect graphs of proper thinness \(2\), the existence of an optimal boundary solution implies the existence of an optimal staggered solution.

\begin{lemma}[Boundary implies staggered]
\label{boundary_implies_staggered}
Let \(G\) be a trivially perfect graph with \(\pthin(G)=2\). If \(G\) admits an optimal boundary solution, then \(G\) also admits an optimal staggered solution.
\end{lemma}
\begin{proof}
    Reversing the order of an optimal boundary solution of $G$ with respect to the universal vertex $u$ yields a staggered solution of $G$, also with respect to $u$: the set $A$ in \cref{def:staggered} is empty, so the class of the solution other than that of $u$, which exists because $\pthin(G)=2$, satisfies both conditions vacuously.
\end{proof}

We now present the main lemmas that characterize the proper thinness of \(\Gplusu\) in terms of the structure of \(G\); they will be combined to obtain the final characterization. The first lemma shows that if \(\pthin(G) = p\) and \(G\) has at least three connected components of proper thinness \(p\), then adding a universal vertex necessarily increases the proper thinness. This holds for arbitrary graphs, not only trivially perfect ones.
A closely related result, due to Bonomo et al.~\cite[Lemma~28]{BGOSS-thin-oper}, states that adding a universal vertex to the disjoint union of three copies of a graph \(H\) yields a graph of proper thinness \(\pthin(H) + 1\). \Cref{prop:pthin-of-three-or-more-max-component} extends it to graphs whose connected components of maximum proper thinness are not necessarily isomorphic, and that may have further connected components of smaller proper thinness.

\begin{lemma}[Three or more components of maximum proper thinness]
\label{prop:pthin-of-three-or-more-max-component}
Let \(G\) be a graph with \(\pthin(G) = p\), for some \(p \ge 1\). If \(G\) has at least three connected components with proper thinness \(p\), then \(\pthin(\Gplusu) = p + 1\).
\end{lemma}

\begin{proof}
Assume, for contradiction, that \(\pthin(\Gplusu) \leq p\), and let \(\Gamma = (S,<)\) be a strongly consistent solution of \(\Gplusu\) with at most \(p\) classes. Let \(C_u\) be the class of \(S\) that contains \(u\), and let \(Comp_1\), \(Comp_2\) and \(Comp_3\) be three connected components of \(G\) of proper thinness \(p\). For each \(j \in \{1,2,3\}\), the restriction of \(\Gamma\) to \(V(Comp_j)\) is a strongly consistent solution of \(Comp_j\) (\cref{obs:restriction-consistent}) with at most as many classes as \(S\). Since \(\pthin(Comp_j) = p\), this restriction uses at least \(p\) classes, so \(S\) has exactly \(p\) classes and every class of \(S\) contains a vertex of \(Comp_j\); in particular, \(C_u\) contains a vertex \(v_j \in V(Comp_j)\).

By \cref{prop:cliques}, the vertices of \(C_u\) smaller than \(u\) form a clique, and so do the vertices of \(C_u\) greater than \(u\); since vertices of different connected components of \(G\) are not adjacent, each of these two cliques is contained in a single connected component of \(G\). Hence the vertices of \(C_u\) other than \(u\) belong to at most two connected components of \(G\), contradicting the fact that \(v_1\), \(v_2\) and \(v_3\) belong to three different ones.

Therefore, \(\pthin(\Gplusu) \ge p + 1\). Since adding a universal vertex can increase the proper thinness by at most one (\cref{thm:join}), it follows that \(\pthin(\Gplusu) = p + 1\).
\end{proof}

We first isolate the construction of a staggered solution of \(\Gplusu\) from one of \(\CompOne\).

\begin{lemma}
\label{lem:staggered-construction}
Let \(G\) be a trivially perfect graph with a connected component \(\CompOne\) of proper thinness \(2\) such that every other connected component of \(G\), if any, has proper thinness \(1\). If \(\CompOne\) admits an optimal staggered solution, then \(\Gplusu\) admits a staggered solution using two classes.
\end{lemma}

\begin{proof}
Let $\Gamma_1=(S_1,<_1)$ be an optimal staggered solution for $\CompOne$ using two classes, let \(u_1\) be the universal vertex of \(\CompOne\) given by the definition of staggered solution (\cref{def:staggered}), and let \(C_{u_1} \in S_1\) be the class containing \(u_1\).
Let $Comp_2,\dots,Comp_k$ be the connected components of $G$ different from $\CompOne$ (possibly none, that is, $k = 1$). Since each $Comp_i$ has proper thinness $1$, for every $i\in\{2,\dots,k\}$ fix a strongly consistent one-class solution $\Gamma_i=(S_i,<_i)$ for $Comp_i$.

Let \(A\) be the set of vertices \(v\) such that \(v \in C_{u_1}\) and \(v < u_1\).
Since $\Gamma_1$ is staggered, there exists a class $C_{sta1}\in S_1$, different from $C_{u_1}$, such that, for every vertex $w\in C_{sta1}$,
\begin{itemize}
    \item $w>A$, and
    \item $w$ is not adjacent to any vertex of $A$.
\end{itemize}
Since \(S_1\) has two classes, \(C_{sta1}\) is the class of \(S_1\) other than \(C_{u_1}\).

We define $\Gamma=(S,<)$ as follows, keeping the order and the class assignment of $\Gamma_1$ on $V(\CompOne)$.
Let \(X =V(Comp_2) \cup \dots \cup V(Comp_k)\). Note that in \(<_1\), every vertex of \(A\) is smaller than every vertex of \(V(\CompOne) \setminus A\): the vertices of \(C_{u_1} \setminus A\) are greater than or equal to \(u_1\), and \(u_1 > A\) by definition of \(A\), while every vertex of the other class \(C_{sta1}\) is greater than every vertex of \(A\) by condition~\hyperlink{prop1_staggered}{(i)} of \cref{def:staggered}. The order \(<\) is obtained from \(<_1\) by placing the new universal vertex \(u\) and the vertices of \(X\) so that every vertex of \(A\) is smaller than \(u\), \(u\) is smaller than every vertex of \(X\), and every vertex of \(X\) is smaller than every vertex of \(V(\CompOne) \setminus A\). We place \(u\) in the class corresponding to \(C_{u_1}\), and we rename this class as \(C_u\).
The components $Comp_2,\dots,Comp_k$ are placed consecutively, in an arbitrary order, and the vertices of each $Comp_i$ are ordered according to $<_i$; in particular, the vertices of each $Comp_i$ are consecutive in $<$. We assign all vertices of $X$ to the class corresponding to $C_{sta1}$, and we denote the resulting class by $C_{sta}$, so that $C_{sta} = C_{sta1} \cup X$. Thus $S = \{C_u, C_{sta}\}$.

By construction, the vertices of \(\CompOne\) smaller than \(u\) are exactly those of \(A\), and
\[
A \;<\; u \;<\; X \;<\; V(\CompOne) \setminus A .
\]
Here \(X\) may be empty. In particular, the vertices of \(X\) lie between \(u\) and \(u_1\).
The construction is illustrated in \cref{fig:staggered-construction}.

The solution \(\Gamma\) satisfies the staggered condition with respect to \(u\). Indeed, the set of vertices of \(C_u\) smaller than \(u\) is \(A\), and every vertex of \(C_{sta}\) is greater than \(u\), hence greater than \(A\): the vertices of \(C_{sta1}\) because they belong to \(V(\CompOne) \setminus A\), and those of \(X\) by construction. Moreover, no vertex of \(C_{sta}\) is adjacent to a vertex of \(A\): the vertices of \(C_{sta1}\) because \(\Gamma_1\) is staggered, and those of \(X\) because they belong to other connected components of \(G\).

\begin{figure}[h]
\centering
\begin{tikzpicture}[x={(0cm,-1cm)}, y={(1cm,0cm)}]

    \begin{scope}
        \node[draw, circle, fill=palCompA, inner sep=2pt] (r0) at (0,3.2) {};

        \node[draw, circle, fill=palCompA, inner sep=2pt,
              label=above:$u_1$] (u1) at (0,2.2) {};

        \node[draw, circle, fill=black, inner sep=2pt,
              label=above:$u$] (u) at (0,-5.8) {};

        \node[draw, circle, fill=palCompA, inner sep=2pt] (v) at (0,-6.5) {};

        \node[draw, circle, fill=palCompA, inner sep=2pt] (r5) at (0,-7.2) {};
              
        \draw[dashed, rounded corners, palClassA, thick]
            (-0.9,-7.5) rectangle (0.9,3.9);

        \node[palClassA] at (0,4.2) {$C_{u}$};
    \end{scope}

    \begin{scope}

        \node[draw, circle, fill=palCompA, inner sep=2pt] (r3) at (3,3.7) {};
        \node[draw, circle, fill=palCompA, inner sep=2pt] (r4) at (3,2.7) {};

        \node[draw, circle, fill=palCompB, inner sep=2pt] (d1) at (3,1.0) {};
        \node[draw, circle, fill=palCompB, inner sep=2pt] (d2) at (3,0.2) {};

        \node[draw, circle, fill=palCompC, inner sep=2pt] (g1) at (3,-0.6) {};
        \node[draw, circle, fill=palCompC, inner sep=2pt] (g2) at (3,-1.6) {};

        \node[draw, circle, fill=palCompE, inner sep=2pt] (b1) at (3,-2.7) {};
        \node[draw, circle, fill=palCompE, inner sep=2pt] (b2) at (3,-3.7) {};

        \node[draw, circle, fill=black, inner sep=1pt] (p1) at (3,-4.6) {};
        \node[draw, circle, fill=black, inner sep=1pt] (p2) at (3,-4.9) {};
        \node[draw, circle, fill=black, inner sep=1pt] (p3) at (3,-5.2) {};

        \draw[dashed, rounded corners, palClassB, thick]
            (2.1,-5.7) rectangle (3.9,4.2);

        \node[palClassB] at (3,4.5) {$C_{sta}$};
    \end{scope}

    \draw (r0) to (u1);
    \draw[bend right=12] (u1) to (v);

    \draw (r3) to (u1);
    \draw (r4) to (u1);
    \draw (r3) to (r0);
   
    \draw[bend right=20] (u) to (r0);

    \draw (b1) to (b2);

    \draw (d1)--(d2);

    \draw [bend right=10] (r5) to (u1);
    \draw  [bend left=20] (r5) to (u);
    \draw (r5) to (v);

    \draw (g1)--(g2);

    \foreach \x in {u1,v,d1,d2,g1,g2,b1,b2, r3, r4} {
        \draw (u)--(\x);
    }

    \node[palCompA!75!black] at (4.8,3.2) {$\CompOne$};
    \node[palCompB!75!black] at (4.8,1.0) {$Comp_2$};
    \node[palCompC!75!black] at (4.8,-1.1) {$Comp_3$};
    \node[palCompE!75!black] at (4.8,-3.5) {$Comp_4$};

\end{tikzpicture}

\caption{
The vertex $u$ is placed in $C_{u}$ and the remaining connected components in the class $C_{sta}$, so that $A < u < X < V(\CompOne) \setminus A$.
}
\label{fig:staggered-construction}
\end{figure}
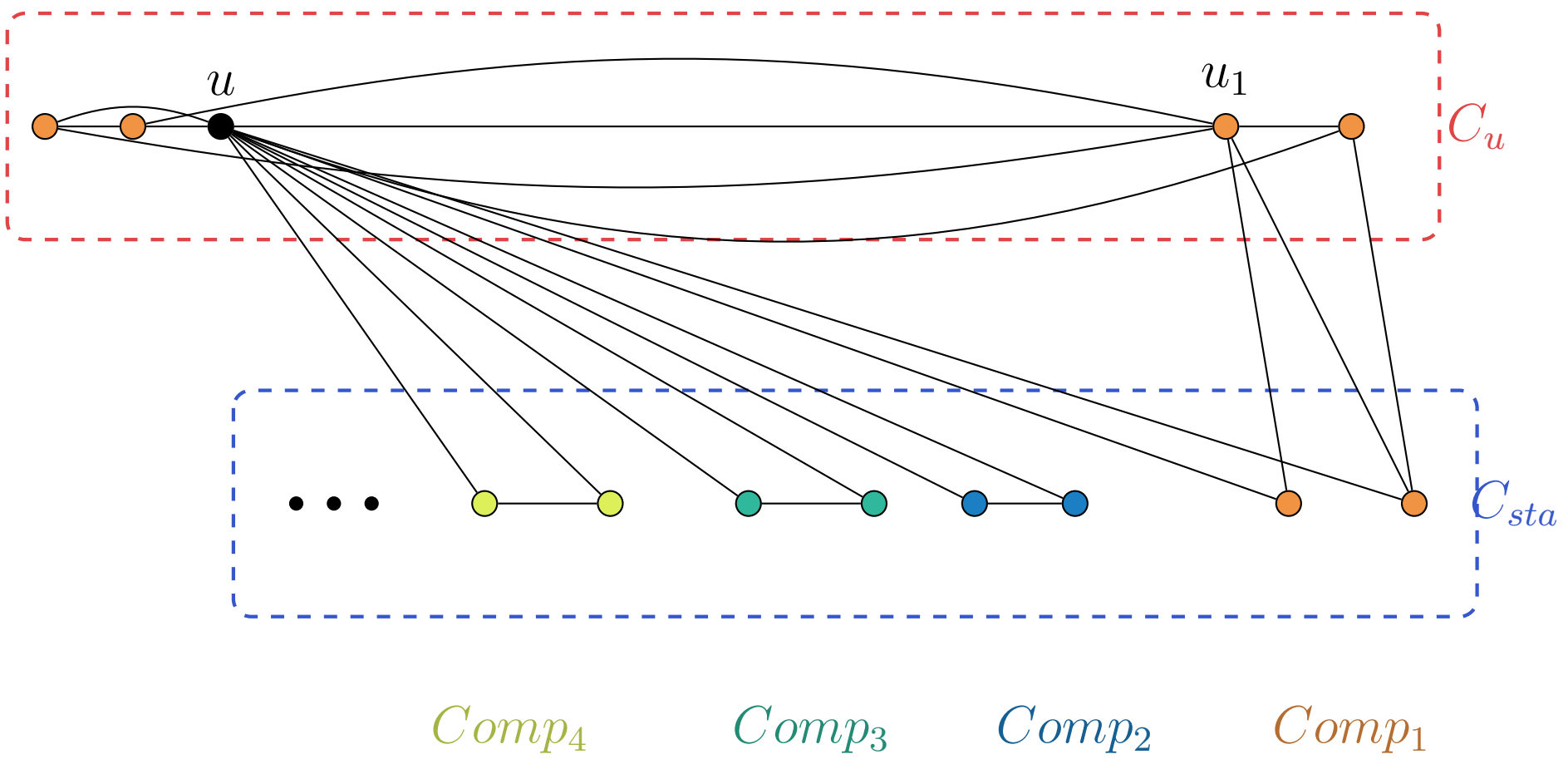

\smallskip
    \noindent\textbf{Strong consistency of $\Gamma$.}
Let $\Gamma_1^{+}$ be the restriction of $\Gamma$ to $V(\CompOne) \cup \{u\}$. It is obtained from $\Gamma_1$ by placing $u$ in the class $C_{u_1}$, smaller than $u_1$ and greater than every vertex of $A$, which are the vertices of $C_{u_1}$ smaller than $u_1$. Since $u_1$ is universal in $\CompOne$, \cref{lem:twin-insertion} shows that $\Gamma_1^{+}$ is a strongly consistent solution of the graph obtained from $\CompOne$ by adding the universal vertex $u$.

Let $v_r < v_s < v_t$ be three vertices of $\Gplusu$. We check conditions~\condi{} and~\condii{} of \cref{def:proper-thinness}, distinguishing three cases.

\emph{The triple does not contain $u$.} The restriction of $\Gamma$ to $V(G)$ coincides with $\Gamma_1$ on $\CompOne$ and with the fixed one-class solution on each $Comp_i$ with $i \ge 2$, so it is strongly consistent on every connected component of $G$. By \cref{lem:components}, if the triple violated condition~\condi{} or~\condii{} of \cref{def:proper-thinness}, then $v_r$ and $v_t$ would belong to the same connected component and $v_s$ to a different one. Since the vertices of each $Comp_i$ with $i \ge 2$ are consecutive in the order, $v_r$ and $v_t$ cannot both lie in such a component; hence $v_r, v_t \in V(\CompOne)$ and $v_s \in X$; in particular, if $X = \emptyset$ no such triple exists. Since $v_r < v_s$, we have $v_r \in A$. Condition~\condi{} does not apply, because $v_r \in C_u$ and $v_s \in C_{sta}$ belong to different classes. For condition~\condii{}, if $v_s$ and $v_t$ belong to the same class, then $v_t \in C_{sta} \cap V(\CompOne) = C_{sta1}$; since $\Gamma_1$ is staggered and $v_r \in A$, no vertex of $C_{sta1}$ is adjacent to $v_r$ (\cref{def:staggered}), so $v_r v_t \notin E(\Gplusu)$ and the condition does not apply either.

\emph{The triple is contained in $V(\CompOne) \cup \{u\}$.} Both conditions hold because $\Gamma_1^{+}$ is strongly consistent.

\emph{The triple contains $u$ and a vertex of $X$.} Every vertex smaller than $u$ belongs to $A$, so $u \neq v_t$. If $u = v_s$, the conclusions $v_s v_t \in E(\Gplusu)$ of~\condi{} and $v_r v_s \in E(\Gplusu)$ of~\condii{} hold because $u$ is universal. If $u = v_r$, the conclusion $v_r v_s \in E(\Gplusu)$ of~\condii{} holds for the same reason, and condition~\condi{} does not apply: the vertices of $C_u$ greater than $u$ are those of $C_{u_1} \setminus A$, which are greater than or equal to $u_1$, and every vertex greater than or equal to $u_1$ belongs to $\CompOne$; so if $v_s$ belonged to $C_u$, then $v_s$ and $v_t$ would both lie in $\CompOne$, and the triple would contain no vertex of $X$.

Therefore, \(\Gamma\) is a strongly consistent solution of \(\Gplusu\) with two classes, and since it satisfies the staggered condition, it is a staggered solution of \(\Gplusu\).
\end{proof}

The next lemma characterizes the case where \(G\) has exactly one component of proper thinness \(2\). In this situation, the proper thinness of \(\Gplusu\) is determined by the existence of an optimal staggered solution for that component.

\begin{lemma}
\label{prop:pthin-2-1-multiple}
Let $G$ be a trivially perfect graph consisting of a connected component $\CompOne$ with proper thinness $2$, and one or more other connected components, each having proper thinness $1$. Let $\Gplusu$ be the graph obtained by adding a universal vertex $u$ to $G$. If $\Gplusu$ has proper thinness 2, then there exists an optimal staggered solution for $\CompOne$.
\end{lemma}
\begin{proof}
Assume that $\Gplusu$ has proper thinness $2$. Let $\Gamma=(S,<)$ be an optimal strongly consistent solution for $\Gplusu$ using two classes, let $C_u$ be the class of $S$ containing $u$, and let $A$ and $B$ be the sets of vertices smaller and greater than $u$, respectively, in $C_u$.

Let $u_1$ be a universal vertex of $\CompOne$, and let $\Gamma_1=(S_1,<_1)$ be the solution of $\CompOne$ given by \cref{lem:transfer} applied to $\Gamma$ and $u_1$. It is strongly consistent with at most two classes, so, since $\pthin(\CompOne)=2$, it is an optimal strongly consistent solution of $\CompOne$ with two classes. We will now show that $\CompOne$ admits an optimal staggered solution.

Note first that $C_u$ contains vertices of $\CompOne$: otherwise every vertex of $\CompOne$ would belong to the other class of $S$, so the restriction of $\Gamma$ to $V(\CompOne)$ would be a strongly consistent solution of $\CompOne$ with a single class (\cref{obs:restriction-consistent}), contradicting $\pthin(\CompOne)=2$.
We distinguish two cases, depending on the positions in \(<\) of the vertices of \(\CompOne\) in \(C_u\) with respect to \(u\).

\smallskip
\noindent\textit{Case 1: every vertex of \(\CompOne\) in \(C_u\) is smaller than \(u\), or every vertex of \(\CompOne\) in \(C_u\) is greater than \(u\).}
By \cref{lem:transfer}, the vertices other than \(u_1\) in the class of \(u_1\) in \(\Gamma_1\) are those of \((C_u \cap V(\CompOne)) \setminus \{u_1\}\), so they are all smaller than \(u_1\) or all greater than \(u_1\). Reversing the order of \(\Gamma_1\) if necessary (\cref{prop:symmetry}), which does not change its partition and hence preserves optimality, we may assume that every vertex of the class of \(u_1\) other than \(u_1\) is greater than \(u_1\). Then the set \(\{x \in C_{u_1} : x <_1 u_1\}\) of \cref{def:staggered} is empty, so the other class of \(\Gamma_1\), which exists because \(\Gamma_1\) has two classes, satisfies both conditions vacuously. Hence \(\Gamma_1\) is an optimal staggered solution for \(\CompOne\).

\smallskip
\noindent\textit{Case 2: \(C_u\) contains a vertex of \(\CompOne\) smaller than \(u\) and a vertex of \(\CompOne\) greater than \(u\).}
By \cref{prop:cliques}, the sets \(A\) and \(B\) are cliques. Each of them contains a vertex of \(\CompOne\), and vertices of different connected components of \(G\) are not adjacent, so every vertex of \(A \cup B\) belongs to \(V(\CompOne)\). Hence \(C_u = A \cup \{u\} \cup B\) contains only \(u\) and vertices of \(\CompOne\).

Let $C_{sta}$ be the other class of $S$. Since $C_u$ contains the vertex $u$ and only vertices from $V(\CompOne)$, every vertex of $G$ not belonging to $V(\CompOne)$ belongs to  $C_{sta}$. Moreover, since $\CompOne$ has proper thinness $2$, the class  $C_{sta}$ must also contain vertices from $V(\CompOne)$. We first locate the vertices outside $\CompOne$ with respect to $A$ and $B$.

\begin{claim} \label{claim_1}
For every vertex \(w\) of \(G\) that does not belong to \(V(\CompOne)\), we have \(B > w > A\).

\end{claim}
\begin{proof}
Suppose there exists a vertex \(t \in B\) such that \(t < s\), where \(s \in V(G) \setminus V(\CompOne)\). Then \(u < t < s\). As $u$ is universal in $G$, the vertices $u$ and $s$ are adjacent. We also have that $u$ and $t$ both belong to $C_u$, and that $ts \notin E(\Gplusu)$, as they belong to different connected components of $G$. Hence $u < t <s$ breaks condition~\condi{} of \cref{def:proper-thinness}, contradicting the strong consistency of $\Gamma$. The case where there exists a vertex \(r \in A\) such that \(r > s\), with \(s \in V(G) \setminus V(\CompOne)\), is analogous.
\end{proof}

Let \(C_{sta}^{1} = C_{sta} \cap V(\CompOne)\) and \(C_{sta}^{\neq 1} = C_{sta} \setminus V(\CompOne)\).

\begin{claim}\label{claim_2}
No vertex of \(C_{sta}^{\neq 1}\) lies between two vertices of \(C_{sta}^{1}\).
\end{claim}

\begin{proof}
 Suppose that there exist vertices $x, x' \in C_{sta}^{1}$ and $s \in C_{sta}^{\neq 1}$ such that $x < s < x'$.
 The vertex $u_1$ is adjacent to both $x$ and $x'$, while it is not adjacent to $s$. As $x$, $s$, and $x'$ all share a class, this breaks either \condi{} or \condii{} of \cref{def:proper-thinness}, depending on if $u_1 < s$ or $s < u_1$.
\end{proof}

The set \(C_{sta}^{\neq 1}\) is not empty, because \(G\) has connected components other than \(\CompOne\); fix \(s_0 \in C_{sta}^{\neq 1}\). By the previous claim, either every vertex of \(C_{sta}^{1}\) is greater than \(s_0\), or every vertex of \(C_{sta}^{1}\) is smaller than \(s_0\). By \cref{prop:symmetry}, reversing \(\Gamma\) and \(\Gamma_1\) if necessary, we may assume the former. \Cref{fig:pthin-2-1-case2} illustrates the resulting situation.

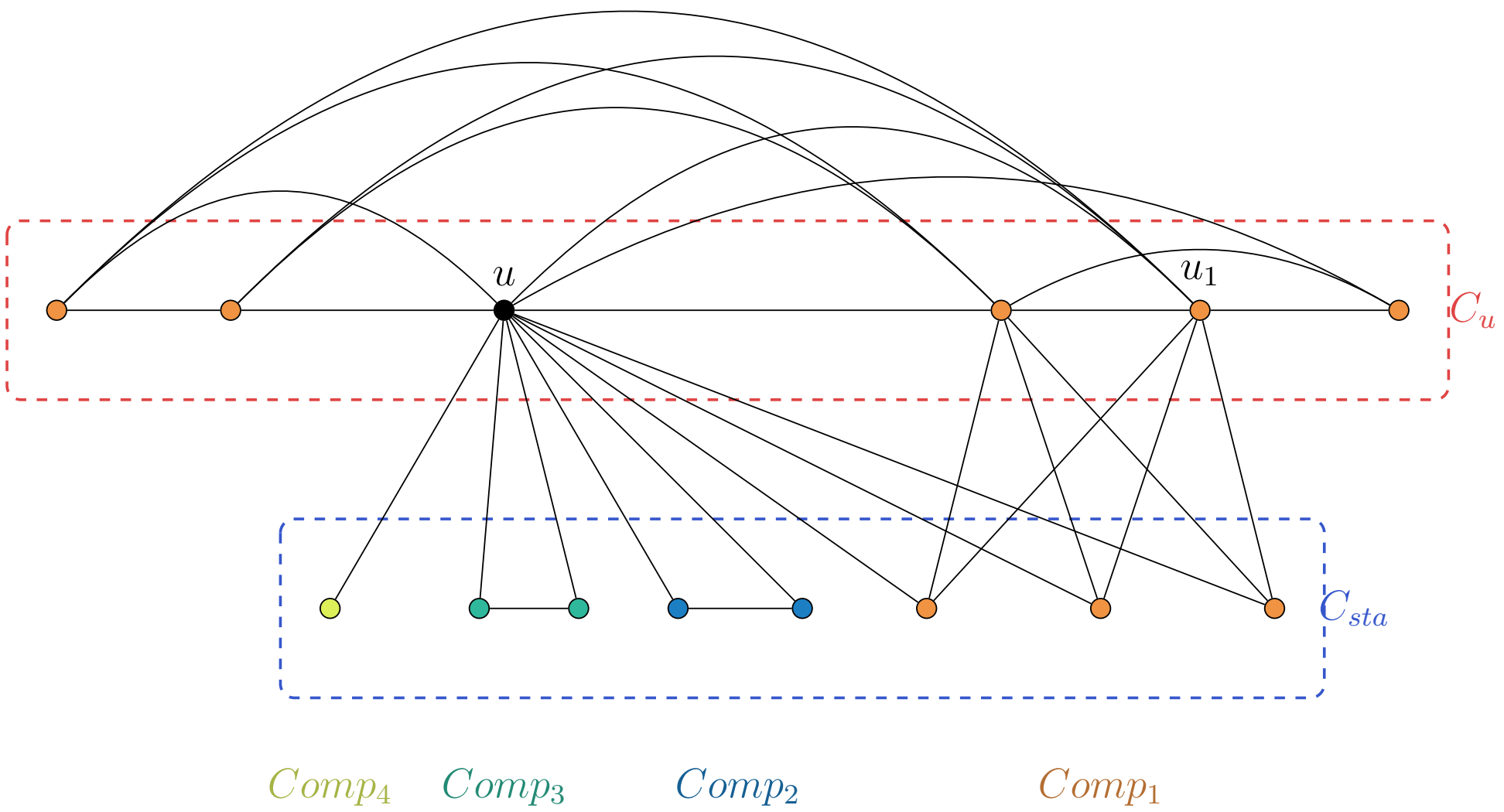
\begin{figure}[ht]
\centering
\scalebox{0.85}{
\begin{tikzpicture}[x={(0cm,-1cm)}, y={(1cm,0cm)}, redv/.style={draw, circle, fill=palCompA, inner sep=2pt},
    orangev/.style={draw, circle, fill=palCompB, inner sep=2pt},
    greenv/.style={draw, circle, fill=palCompC, inner sep=2pt},
    bluev/.style={draw, circle, fill=palCompE, inner sep=2pt},
    blackv/.style={draw, circle, fill=black, inner sep=2pt}]

    \begin{scope}
        \node[redv, label=above:$u_1$] (r0) at (0,5.25) {};
        \node[redv] (v) at (0,3.25) {};
        \node[blackv, label=above:$u$] (u) at (0,-1.75) {};
        \node[redv] (v1) at (0,-4.5) {};
        \node[redv] (r5) at (0,-6.25) {};

        \node[redv] (r11) at (0,7.25) {};

        \draw[dashed, rounded corners, palClassA, thick]
            (-0.9,-6.75) rectangle (0.9,7.75);

        \node[palClassA] at (0,8) {$C_{u}$};
    \end{scope}

    \begin{scope}
        \node[redv] (r8) at (3,6) {};
        \node[redv] (r9) at (3,4.25) {};
        \node[redv] (r10) at (3,2.5) {};

        \node[orangev] (d1) at (3,1.25) {};
        \node[orangev] (d2) at (3,0) {};

        \node[greenv] (g1) at (3,-1) {};
        \node[greenv] (g2) at (3,-2) {};

        \node[bluev] (b1) at (3,-3.5) {};

        \draw[dashed, rounded corners, palClassB, thick]
            (2.1,-4.0) rectangle (3.9,6.5);

        \node[palClassB] at (3,6.8) {$C_{sta}$};
    \end{scope}

    \draw (v1) to (u);
    \draw [in=180, out=0] (u) to (v);
    \draw (v) to (r0);
    \draw [bend left=45, looseness=1.25] (u) to (r0);
    \draw (v1) to (r5);
    \draw [bend right=45, looseness=1.25] (u) to (r5);

    \draw (u) to (g2);
    \draw (u) to (d2);
    \draw (u) to (d1);
    \draw (u) to (r10);
    \draw (u) to (r9);
    \draw (u) to (r8);
    \draw (u) to (g1);
    \draw (u) to (b1);
    \draw [bend left=30, looseness=1](u) to (r11);

    \draw [bend right=30, looseness=1](r11) to (v);
    \draw(r11) to (r0);

    \draw (d1) to (d2);
    \draw (v) to (r8);
    \draw (v) to (r9);
    \draw (r0) to (r8);
    \draw [bend right=45, looseness=1.25](r0) to (v1);
    \draw [bend right=45, looseness=1.25](r0) to (r5);
    \draw (r0) to (r9);
    \draw (r0) to (r10);
    \draw (v) to (r10);
    \draw [bend right=45, looseness=1.25](v) to (v1);
    \draw [bend right=45, looseness=1.25](v) to (r5);

    \draw (g1) to (g2);

    \node[palCompA!75!black] at (4.8,4.25) {$\CompOne$};
    \node[palCompB!75!black] at (4.8,0.6) {$Comp_2$};
    \node[palCompC!75!black] at (4.8,-1.75) {$Comp_3$};
    \node[palCompE!75!black] at (4.8,-3.5) {$Comp_4$};

\end{tikzpicture}
}

\caption{
The vertex $u$ is placed in $C_u$ with $A<u<B$, while the remaining connected components are placed in the class $C_{sta}$ between $A$ and $B$. In the example, the vertices of \(\CompOne\) in \(C_{sta}\) are greater than the other vertices of \(C_{sta}\).
}
\label{fig:pthin-2-1-case2}
\end{figure}

We claim that \(\Gamma_1\) is staggered with respect to \(u_1\), witnessed by its class coming from \(C_{sta}\), which is contained in \(C_{sta}^{1}\) by \cref{lem:transfer}. This class is different from the class of \(u_1\) in \(\Gamma_1\), which comes from \(C_u\), and the vertices of the class of \(u_1\) smaller than \(u_1\) belong to \(A\). Hence it suffices to show that every vertex of \(C_{sta}^{1}\) is greater than every vertex of \(A\) and is not adjacent to any of them.

Let \(w \in C_{sta}^{1}\) and \(x \in A\). By \cref{claim_1} and the choice of \(s_0\), we have \(x < s_0 < w\), so condition~\hyperlink{prop1_staggered}{(i)} holds. If \(xw \in E(\CompOne)\), the triple \(x < s_0 < w\) would contradict the strong consistency of \(\Gamma\), since \(s_0\) and \(w\) belong to the same class and \(x s_0 \notin E(\Gplusu)\); hence condition~\hyperlink{prop2_staggered}{(ii)} holds as well.

Therefore, in all cases, there exists an optimal staggered solution for $\CompOne$.
\end{proof}

The following lemma addresses the case where \(G\) has exactly two components of proper thinness \(2\): then the existence of a strongly consistent solution in two classes for \(\Gplusu\) depends solely on the existence of optimal boundary solutions for both components.

\begin{lemma}[Two components of proper thinness 2]
\label{prop:pthin-of-two-max-component}
Let $G$ be a trivially perfect graph with proper thinness $2$, consisting of two connected components $\CompOne$ and $Comp_2$ of proper thinness $2$, and any number of additional connected components of proper thinness $1$.

Then $\Gplusu$ has proper thinness $2$ if and only if there exists an optimal boundary solution for $\CompOne$, and there exists an optimal boundary solution for $Comp_2$.
\end{lemma}

\begin{proof}
\noindent\textbf{(\(\Rightarrow\))} Assume that $\Gplusu$ has proper thinness $2$, and let $\Gamma=(S,<)$ be an optimal strongly consistent solution for $\Gplusu$ where $S = \{C_u, C\}$ and $C_u$ is the class containing the universal vertex $u$. Since both $\CompOne$ and $Comp_2$ have proper thinness $2$, the class $C_u$ must contain vertices from both $\CompOne$ and $Comp_2$.

By \cref{prop:cliques}, all vertices of $C_u$ smaller than $u$ form a clique, and all vertices of $C_u$ greater than $u$ form a clique. Since vertices of $\CompOne$ and $Comp_2$ are not adjacent, either every vertex of $\CompOne$ in $C_u$ is greater than $u$ and every vertex of $Comp_2$ in $C_u$ is smaller than $u$, or every vertex of $\CompOne$ in $C_u$ is smaller than $u$ and every vertex of $Comp_2$ in $C_u$ is greater than $u$. Without loss of generality, by \cref{prop:symmetry}, assume that all vertices of $\CompOne$ in $C_u$ are greater than $u$, and all vertices of $Comp_2$ in $C_u$ are smaller than $u$.

For each \(i \in \{1,2\}\), let \(u_i\) be a universal vertex of \(Comp_i\), and let \(\Gamma_i\) be the solution of \(Comp_i\) given by \cref{lem:transfer} applied to \(\Gamma\) and \(u_i\). It is strongly consistent with at most two classes, hence optimal because \(\pthin(Comp_i)=2\), and the other vertices of the class of \(u_i\) are those of \((C_u \cap V(Comp_i)) \setminus \{u_i\}\), each smaller than \(u_i\) if and only if it is smaller than \(u\). Hence \(u_2\) is the greatest vertex of its class in \(\Gamma_2\), so \(\Gamma_2\) is an optimal boundary solution of \(Comp_2\), and \(u_1\) is the smallest vertex of its class in \(\Gamma_1\), so reversing the order of \(\Gamma_1\) (\cref{prop:symmetry}) yields an optimal boundary solution of \(\CompOne\).

\noindent\textbf{(\(\Leftarrow\))} Assume that there exists an optimal boundary solution \(\Gamma_1=(S_1,<_1)\) for \(\CompOne\) with respect to a universal vertex \(u_1\), and an optimal boundary solution \(\Gamma_2=(S_2,<_2)\) for \(Comp_2\) with respect to a universal vertex \(u_2\). For each \(i \in \{1,2\}\), the partition is \(S_i = \{C_{u_i}, C_i\}\), where \(C_{u_i}\) is the class containing the universal vertex \(u_i\). By \cref{prop:symmetry}, we may assume that $u_1$ is the greatest vertex of $C_{u_1}$ in $\Gamma_1$, and that $u_2$ is the smallest vertex of $C_{u_2}$ in $\Gamma_2$. Since $\CompOne$ and $Comp_2$ both have proper thinness $2$, we have $\pthin(\Gplusu)\ge 2$ by \cref{thm:hereditary}.

Let \(Comp_3,\dots,Comp_k\) be the connected components of \(G\) different from \(\CompOne\) and \(Comp_2\). Since each of these components has proper thinness \(1\), we fix a strongly consistent one-class solution \(\Gamma_i = (S_i, <_i)\) for each \(i \in \{3,\dots,k\}\).
We construct a solution $\Gamma=(S,<)$ for $\Gplusu$ using two classes as follows. We define the classes \(C_u = C_{u_1} \cup C_{u_2} \cup \{u\}\) and \(C = C_1 \cup C_2 \cup V(Comp_3) \cup \dots \cup V(Comp_k)\). In the order \(<\), every vertex of \(\CompOne\) is smaller than \(u\), and every vertex of \(Comp_3 \cup \dots \cup Comp_k \cup Comp_2\) is greater than \(u\).
For every \(i \in \{3,\dots,k\}\), we place every vertex of \(Comp_i\) greater than every vertex of \(\CompOne\) and smaller than every vertex of \(Comp_2\). Within each \(Comp_i\), we respect the order of the fixed strongly consistent one-class solution \(\Gamma_i\). The connected components \(Comp_3,\dots,Comp_k\) are placed in an arbitrary fixed order between the vertices of \(\CompOne\) and the vertices of \(Comp_2\).
For each connected component of \(G\), we keep the relative order of its vertices as given by its corresponding strongly consistent solution \(\Gamma_i\) for \(i \in \{1,\dots,k\}\). Moreover, the vertices of each connected component appear consecutively in the ordering \(<\).
Thus \(V(\CompOne) < u < V(Comp_3) \cup \dots \cup V(Comp_k) < V(Comp_2)\), and \(u_1\) and \(u_2\) are, respectively, the greatest vertex of \(C_u\) smaller than \(u\) and the smallest vertex of \(C_u\) greater than \(u\).

\noindent\textbf{Strong consistency of $\Gamma$.}
We use \cref{lem:universal-insertion} for the universal vertex $u$. The restriction of $\Gamma$ to $V(G)$ coincides with $\Gamma_i$ on each $Comp_i$, and the vertices of each component are consecutive, so it is strongly consistent by \cref{cor:contiguous-components}. The vertices of $C_u$ greater than $u$ are those of $C_{u_2}$, whose smallest vertex is $u_2$, and every vertex greater than a vertex of $Comp_2$ belongs to $Comp_2$; hence \cref{cond:ins-above} for $u$ follows from \cref{cond:ins-above} for $u_2$ in $\Gamma_2$, together with the fact that $u_2$ is universal in $Comp_2$. Symmetrically (the vertices of $C_u$ smaller than $u$ are those of $C_{u_1}$, whose greatest vertex is $u_1$, and every vertex smaller than a vertex of $\CompOne$ belongs to $\CompOne$), \cref{cond:ins-below} for $u$ follows from \cref{cond:ins-below} for $u_1$ in $\Gamma_1$. Therefore, $\Gamma$ is a strongly consistent solution of $\Gplusu$ with two classes, and hence $\pthin(\Gplusu)=2$.
\end{proof}

We now turn to staggered solutions. The following lemma shows that if \(G\) has two components of proper thinness \(2\) and \(\pthin(\Gplusu)=2\), then \(\Gplusu\) admits no optimal staggered solution.

\begin{lemma}[Two components exclude staggered solutions]\label{lem:two_components_no_staggered}
Let \( G \) be a trivially perfect graph with proper thinness \(2\) such that two of its connected components, \( \CompOne \) and \( Comp_2 \), have proper thinness \(2\). Suppose that \( \Gplusu \) has proper thinness \(2\). Then \( \Gplusu \) does not admit an optimal staggered solution.
\end{lemma}

\begin{proof}
Assume, for the sake of contradiction, that $\Gplusu$ admits an optimal staggered solution $\Gamma = (S,<)$ with two classes. Write $S = \{C_u, Z\}$ where $C_u$ is the class containing $u$. We may assume that $u$ is the universal vertex that witnesses the staggered condition. Since $\pthin(\CompOne)=\pthin(Comp_2)=2$, the restriction of $\Gamma$ to each of $\CompOne$ and $Comp_2$ must use both classes.

Let $A = \{x \in C_u : x < u\}$ and $B = \{x \in C_u : x > u\}$. By \cref{prop:cliques}, $A$ and $B$ are cliques, so each of them is contained in a single connected component of $G$. Since $\CompOne$ and $Comp_2$ both have vertices in $C_u \setminus \{u\} = A \cup B$, one of them contains $A$ and the other one contains $B$; we may assume that $A \subseteq V(\CompOne)$. Because $\CompOne$ has proper thinness $2$, it must contain a vertex $w \in Z$. By \cref{def:staggered}, the class that witnesses the staggered condition is different from $C_u$, so it is $Z$. Hence, $w > A$ and $w$ is not adjacent to any vertex in \(A\).

Since $\CompOne$ is a connected trivially perfect graph, it contains at least one universal vertex $u_1$~\cite{Brandstadt-book}. This vertex $u_1$ is either in $A$ or in $Z$. In both cases, this yields an edge between a vertex in $A$ and $Z \cap V(\CompOne)$, contradicting that $\Gamma$ is staggered.
\end{proof}

We also need a criterion for the existence of optimal boundary solutions of \(\Gplusu\), given by the following lemma.

\begin{lemma}[Characterization of boundary solutions]\label{lemma:optimal_boundary_solution}
Let \(G\) be a trivially perfect graph, and suppose $\Gplusu$ has proper thinness 2. Then $\Gplusu$ admits an optimal boundary solution if and only if either \(\pthin(G) = 1\), or \(G\) has exactly one connected component of proper thinness \(2\) that itself admits an optimal boundary solution.
\end{lemma}

\begin{proof}
\noindent\textbf{(\(\Rightarrow\))} Let \(\Gamma\) be an optimal boundary solution of \(\Gplusu\). Since \(\Gplusu\) has proper thinness \(2\), \(\Gamma\) has exactly two classes: \(C_u\), which contains the universal vertex \(u\), and \(C_1\), the other class. As \(u\) and the vertex witnessing that \(\Gamma\) is boundary are both universal, exchanging them is an automorphism of \(\Gplusu\), so we may assume that \(\Gamma\) is boundary with respect to \(u\).

By \cref{boundary_implies_staggered}, \(\Gplusu\) admits an optimal staggered solution. No component of \(G\) has proper thinness at least \(3\), by \cref{thm:hereditary}, since \(\pthin(\Gplusu)=2\). If \(G\) had two components of proper thinness \(2\) (three are excluded by \cref{prop:pthin-of-three-or-more-max-component}), then \(\pthin(G)=2\) by \cref{thm:hereditary}, contradicting \cref{lem:two_components_no_staggered}.

If \(G\) has exactly one such component \(\CompOne\), let \(u_1\) be a universal vertex of it and \(\Gamma_1\) be given by \cref{lem:transfer} applied to \(\Gamma\) and \(u_1\). It is strongly consistent with at most two classes, hence optimal as \(\pthin(\CompOne)=2\), and \(u_1\) is the greatest vertex of its class, since \(u\) is; so \(\Gamma_1\) is an optimal boundary solution of \(\CompOne\).

Finally, if \(G\) had no connected component of proper thinness \(2\), then all its components would have proper thinness \(1\), and hence, by \cref{thm:union}, \(G\) would have proper thinness \(1\).

\noindent\textbf{(\(\Leftarrow\))} First, suppose \(G\) has a unique connected component \(\CompOne\) of proper thinness \(2\). Assume further that \(\CompOne\) admits an optimal boundary solution \(\Gamma_1 = (S_1, <_1)\), with partition \(S_1 = \{C_{u_1}, C_1\}\). The remaining components of \(G\), namely \(Comp_2, \dots, Comp_m\) (possibly none), have proper thinness \(1\). For each \(i > 1\), let \(\Gamma_i = (S_i, <_i)\) be a strongly consistent solution of \(Comp_i\) with one class, which exists because \(\pthin(Comp_i) = 1\). We now show that \(\Gplusu\) admits an optimal boundary solution. Let $u_1 \in \CompOne$ be the universal vertex that is greatest in its class $C_{u_1}$, following the definition of boundary solution (\cref{def:boundary-solution}).

We now insert the components \(Comp_2, \dots, Comp_m\) into the class \(C_1\). We fix an arbitrary order \(Comp_2, \dots, Comp_m\) of these components and place their vertices so that \(V(\CompOne) < V(Comp_2) < \dots < V(Comp_m)\), and we order the vertices of each \(Comp_i\) according to \(<_i\); in particular, the vertices of each \(Comp_i\) are consecutive in the order. The resulting class is denoted by \(C\).
Next, we add the universal vertex \(u\) to the class \(C_{u_1}\). We set \(u > u_1\), so that \(u\) becomes the greatest vertex of the new class \(C_u\) and so that there is no vertex \(w\) in \(\Gplusu\) such that \(u_1 < w < u\). Let \(\Gamma = (S,<)\) be the resulting solution of \(\Gplusu\), where \(S = \{C_u, C\}\). Thus \(V(\CompOne) \cup \{u\} < V(Comp_2) < \dots < V(Comp_m)\).

\noindent\textbf{Strong consistency of \(\Gamma\).}
We check the triples of \(\Gplusu\) in three groups. By \cref{lem:twin-insertion}, applied to \(\Gamma_1\) and \(u_1\), the restriction of \(\Gamma\) to \(V(\CompOne) \cup \{u\}\) is strongly consistent, which covers the triples contained in this set. The restriction of \(\Gamma\) to \(V(G)\) is strongly consistent by \cref{cor:contiguous-components}, since it coincides with \(\Gamma_i\) on each \(Comp_i\) and the vertices of each component are consecutive, which covers the triples not containing \(u\). In the remaining triples, which contain \(u\) and a vertex of some \(Comp_i\) with \(i \ge 2\), the vertex \(u\) is not the greatest one, because every vertex of \(Comp_i\) is greater than \(u_1\) and no vertex lies between \(u_1\) and \(u\); the conclusions of conditions~\condi{} and~\condii{} that involve \(u\) hold because \(u\) is universal, and if \(u\) is the smallest vertex of the triple, condition~\condi{} does not apply, because \(u\) is the greatest vertex of \(C_u\). Therefore, \(\Gamma\) is a strongly consistent solution of \(\Gplusu\) with two classes, and by construction of \(C_u\) it satisfies the boundary condition (\cref{def:boundary-solution}); since \(\pthin(\Gplusu)=2\), \(\Gamma\) is an optimal boundary solution.

We now consider the case where \(\pthin(G)=1\) and adding a universal vertex \(u\) increases the proper thinness. Let \(\Gamma' = (S', <')\) be a strongly consistent solution of \(G\), with \(S' = \{C\}\). We define a solution \(\Gamma = (S, <)\) for \(\Gplusu\) by setting \(S = \{C, C_u\}\), where \(C_u = \{u\}\), and letting \(<\) be obtained from \(<'\) by placing \(u\) at an arbitrary position. By \cref{lem:universal-insertion}, the solution $\Gamma$ is strongly consistent, since \(C_u = \{u\}\), so only \cref{cond:ins-restriction} has to be checked, and it holds because the restriction of \(\Gamma\) to \(V(G)\) is \(\Gamma'\).
Furthermore, by construction of \(C_u\), the solution \(\Gamma\) satisfies the boundary condition (\cref{def:boundary-solution}), and since \(\pthin(\Gplusu)=2\), it is an optimal boundary solution.
\end{proof}

\subsection{A linear-time algorithm}

Let $G$ be a trivially perfect graph with $n$ vertices and $m$ edges, given by adjacency lists. We describe an algorithm that computes $\min(\pthin(G),3)$ in time $O(n+m)$, and in particular decides whether $\pthin(G)\le2$. Since $G$ is built from single vertices by disjoint unions and universal vertex additions (\cref{def:triviallyPerfect-recursive}), the algorithm recovers the sequence of operations that builds $G$ and evaluates along it the recursion given by the results of the previous subsection.

We record these operations in a tree. Consecutive universal vertex additions produce graphs of the same proper thinness (\cref{prop:many_universal_vertices}), so we group them into a single node, which adds a set of universal vertices.

\begin{definition}\label{def:tree}
  The \emph{tree of operations} $T(G)$ of a trivially perfect graph $G$ is the
  rooted tree whose nodes are the operations of \cref{def:triviallyPerfect-recursive} that
  build $G$, each node having as children the operations that produce its
  arguments, and such that
  \begin{enumerate}
    \item the arguments of every disjoint union are connected graphs, and there
      are at least two of them;
    \item the argument of every universal vertex addition is a disjoint union,
      and the set $U$ of vertices that it adds is the whole set of universal
      vertices of the graph it produces; and
    \item the leaves are complete graphs: a leaf $K_n$ stands for a $K_1$
      followed by the addition of $n-1$ universal vertices.
  \end{enumerate}
  We denote by $G_x$ the graph produced at the node $x$, and we call $x$ a
  \emph{union node} or an \emph{extension node} according to the operation it
  performs.
\end{definition}

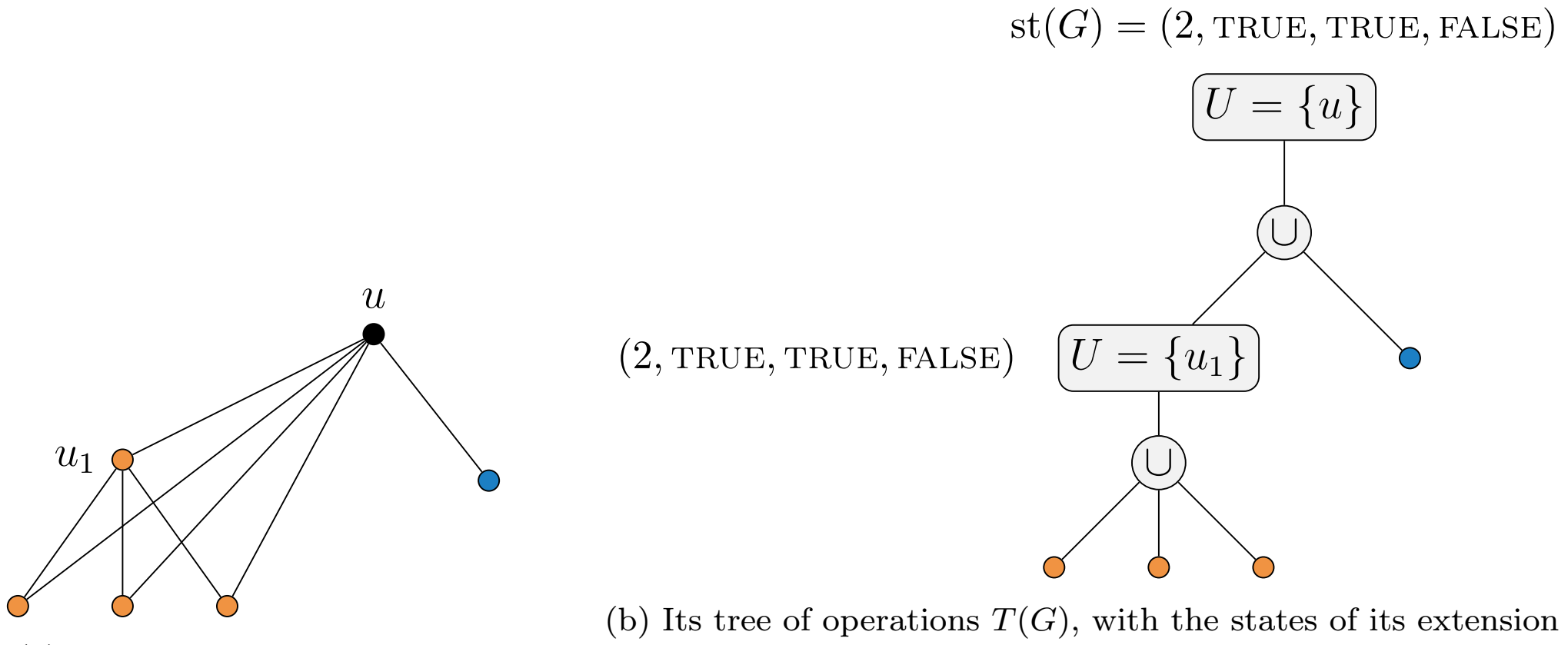
\begin{figure}[ht]
\centering
\begin{subfigure}[b]{0.34\textwidth}
\centering
\scalebox{0.9}{
\begin{tikzpicture}[
    orangev/.style={draw, circle, fill=palCompA, inner sep=2pt},
    lightbluev/.style={draw, circle, fill=palCompB, inner sep=2pt},
    blackv/.style={draw, circle, fill=black, inner sep=2pt},
]
    \node[blackv, label=above:$u$] (u) at (1.2,3.0) {};
    \node[orangev, label=left:$u_1$] (u1) at (-1.2,1.8) {};
    \node[orangev] (l1) at (-2.2,0.4) {};
    \node[orangev] (l2) at (-1.2,0.4) {};
    \node[orangev] (l3) at (-0.2,0.4) {};
    \node[lightbluev] (z) at (2.3,1.6) {};
    \draw (u) to (u1);
    \draw (u) to (z);
    \draw (u) to (l1);
    \draw (u) to (l2);
    \draw (u) to (l3);
    \draw (u1) to (l1);
    \draw (u1) to (l2);
    \draw (u1) to (l3);
\end{tikzpicture}
}
\caption{A trivially perfect graph \(G\).}
\label{fig:tree-a}
\end{subfigure}
\hfill
\begin{subfigure}[b]{0.62\textwidth}
\centering
\scalebox{0.9}{
\begin{tikzpicture}[
    ext/.style={draw, rounded corners, inner sep=3pt, fill=gray!10},
    union/.style={draw, circle, inner sep=1.5pt, fill=gray!10},
    orangev/.style={draw, circle, fill=palCompA, inner sep=2pt},
    lightbluev/.style={draw, circle, fill=palCompB, inner sep=2pt},
    state/.style={font=\small},
]
    \node[ext] (e1) at (0,4.2) {$U = \{u\}$};
    \node[union] (n1) at (0,3.0) {$\cup$};
    \node[ext] (e2) at (-1.2,1.8) {$U = \{u_1\}$};
    \node[lightbluev] (z) at (1.2,1.8) {};
    \node[union] (n2) at (-1.2,0.8) {$\cup$};
    \node[orangev] (l1) at (-2.2,-0.2) {};
    \node[orangev] (l2) at (-1.2,-0.2) {};
    \node[orangev] (l3) at (-0.2,-0.2) {};
    \draw (e1) to (n1);
    \draw (n1) to (e2);
    \draw (n1) to (z);
    \draw (e2) to (n2);
    \draw (n2) to (l1);
    \draw (n2) to (l2);
    \draw (n2) to (l3);

    \node[state, anchor=south, yshift=0.1cm] at (e1.north) {$\mathrm{st}(G) = (2, \textsc{true}, \textsc{true}, \textsc{false})$};
    \node[state, anchor=east, xshift=-0.25cm] at (e2.west) {$(2, \textsc{true}, \textsc{true}, \textsc{false})$};
\end{tikzpicture}
}
\caption{Its tree of operations \(T(G)\), with the states of its extension nodes.}
\label{fig:tree-b}
\end{subfigure}
\caption{(a)~The graph of \cref{fig:stag-bnd}. (b)~Its tree of operations. Each extension node has its state (\cref{def:state}).}
\label{fig:tree}
\end{figure}

With this normalization, $T(G)$ is unique up to the ordering of the children of
the union nodes, it has at most $2n$ nodes, and the graph $G_x$ is connected if
and only if $x$ is a leaf or an extension node. \Cref{fig:tree} shows a trivially perfect graph together with its tree of operations.

The recognition algorithm of \citet{YanChenChang} computes in time $O(n+m)$ a rooted forest in which two vertices are ancestor and descendant if and only if they are adjacent in $G$, and the tree of operations is read off from it in time $O(n)$.

\begin{theorem}[{\cite{YanChenChang}}]\label{thm:tree-linear}
  The tree of operations of a trivially perfect graph with $n$ vertices and $m$
  edges can be computed in time $O(n+m)$.
\end{theorem}

Besides the results of the previous subsection and
\cref{thm:union,thm:join,thm:hereditary}, we must distinguish proper thinness $1$ from $2$, that is, by \cref{lem:claw-free-pthin1}, decide whether the graph is claw-free; the following lemma turns this into a condition on the connected components
that the recursion already has at hand.

\begin{lemma}\label{lem:clawfree}
  Let $H$ be a connected trivially perfect graph, let $U$ be its set of universal
  vertices and let $H_1,\dots,H_k$ be the connected components of $H-U$. Then $H$
  is claw-free if and only if $k=0$, or $k=2$ and both $H_1$ and $H_2$ are
  complete graphs.
\end{lemma}

\begin{proof}
  Note first that $k\ne1$, since otherwise a universal vertex of the connected trivially perfect graph $H_1$ would be universal in $H$. If $k=0$, then $H$ is complete, hence claw-free. If $k\ge3$, or if $k=2$ and, say, $H_1$ is not complete, then a vertex of $U$ together with three pairwise non-adjacent vertices, namely one vertex of each of $H_1,H_2,H_3$ in the first case, and two non-adjacent vertices of $H_1$ and a vertex of $H_2$ in the second, induces a claw. Conversely, if $k=2$ and $H_1$ and $H_2$ are complete, then a set of at least two pairwise non-adjacent vertices of $H$ contains no vertex of $U$ and at most one vertex of each $H_i$, so $H$ has no three pairwise non-adjacent vertices and hence no induced claw.
\end{proof}

We can now describe what the recursion computes at each node of the tree of
operations.

\begin{definition}\label{def:state}
  Let $H$ be a connected trivially perfect graph. The \emph{state} of $H$ is the
  tuple
  \[
    \mathrm{st}(H)=\bigl(p(H),\ \mathit{stag}(H),\ \mathit{bnd}(H),\
    \mathit{complete}(H)\bigr),
  \]
  where $p(H)=\min(\pthin(H),3)$, where $\mathit{complete}(H)$ indicates
  whether $H$ is a complete graph and where, provided that $p(H)=2$, the values
  $\mathit{stag}(H)$ and $\mathit{bnd}(H)$ indicate whether $H$ admits an optimal
  staggered solution (\cref{def:staggered}) and whether it admits
  an optimal boundary solution (\cref{def:boundary-solution}),
  respectively. If $p(H)\ne2$, both values are set to \textsc{false}.
\end{definition}

The next proposition, the recursion evaluated by the algorithm, determines the state of a connected trivially perfect graph from the states of the connected components that remain after removing its universal vertices.

\begin{proposition}\label{prop:dp}
  Let $H$ be a connected trivially perfect graph.
  \begin{enumerate}
    \item\label{item:Kn} If $H=K_n$, then
      $\mathrm{st}(H)=(1,\textsc{false},\textsc{false},\textsc{true})$.
    \item\label{item:not-Kn} Let $H$ be obtained from $D=Comp_1\cup\dots\cup Comp_k$
      with $k\ge2$, where $Comp_1,\dots,Comp_k$ are the connected components of $D$,
      by adding a set $U$ of universal vertices, and let $c_2$ be the
      number of indices $i$ with $p(Comp_i)=2$. Then $H$ is not a
      complete graph, so $\mathit{complete}(H)=\textsc{false}$, and:
      \begin{enumerate}
        \item if $p(Comp_i) = 3$ for some $i$, then
          $\mathrm{st}(H)=(3,\textsc{false},\textsc{false},\textsc{false})$;
        \item otherwise, if $c_2=0$, then
          $\mathrm{st}(H)=(1,\textsc{false},\textsc{false},\textsc{false})$ when
          $k=2$ and both $Comp_1$ and $Comp_2$ are complete
          graphs, and
          $\mathrm{st}(H)=(2,\textsc{true},\textsc{true},\textsc{false})$
          otherwise;
        \item otherwise, if $c_2=1$ and $Comp_i$ is the component of
          proper thinness $2$, then $\mathrm{st}(H)=\bigl(2,\textsc{true},
          \mathit{bnd}(Comp_i),\textsc{false}\bigr)$ if
          $\mathit{stag}(Comp_i)$ holds, and
          $\mathrm{st}(H)=(3,\textsc{false},\textsc{false},\textsc{false})$
          otherwise;
        \item otherwise, if $c_2=2$ and $Comp_i,Comp_j$ are the
          components of proper thinness $2$, then
          $\mathrm{st}(H)=(2,\allowbreak\textsc{false},\allowbreak\textsc{false},\allowbreak\textsc{false})$ if
          $\mathit{bnd}(Comp_i)$ and $\mathit{bnd}(Comp_j)$
          both hold, and
          $\mathrm{st}(H)=(3,\textsc{false},\textsc{false},\textsc{false})$
          otherwise;
        \item otherwise $c_2\ge3$, and
          $\mathrm{st}(H)=(3,\textsc{false},\textsc{false},\textsc{false})$.
      \end{enumerate}
  \end{enumerate}
\end{proposition}

\begin{proof}
  \Cref{item:Kn} is clear. For \cref{item:not-Kn}, note that, since $D$ is disconnected, $U$ is the set of all universal vertices of $H$, and $H-U=D$. Let $u \in U$, and let $u' \in U \setminus \{u\}$, if any. The graph $H-u'$ is a connected trivially perfect graph, since it is an induced subgraph of $H$ and $u$ is universal in it, so $\pthin(H-u') = \pthin(H)$ by \cref{prop:many_universal_vertices}. If this value is $2$, then $H-u'$ is its own unique connected component of proper thinness $2$, and \cref{lem:boundary-and-staggered-preserved}, applied to $H-u'$ and the universal vertex $u'$, shows that $H-u'$ and $H$ have the same values of $\mathit{stag}$ and $\mathit{bnd}$; otherwise both values are \textsc{false} for both graphs by \cref{def:state}. Repeating this, we may assume that $U=\{u\}$. The graph $H$ is not complete, since
  the different components $Comp_1$ and $Comp_2$ of $H-U$ contain non-adjacent vertices, and
  $\pthin(D)=\max_i\pthin(Comp_i)$ by
  \cref{thm:union}.

  In case (a), $\pthin(H)\ge\pthin(Comp_i)\ge3$ by
  \cref{thm:hereditary}. In all the remaining cases,
  $\pthin(Comp_i)\le2$ for every $i$. In case (b) we have $\pthin(D)=1$, so $\pthin(H)\le2$ by
  \cref{thm:join}. By \cref{lem:claw-free-pthin1} and \cref{lem:clawfree},
  $\pthin(H)=1$ if and only if $k=2$ and both $Comp_1$ and
  $Comp_2$ are complete graphs. Otherwise $\pthin(H)=2$, and $H$
  admits an optimal boundary solution by \cref{lemma:optimal_boundary_solution}, and hence also an
  optimal staggered solution by \cref{boundary_implies_staggered}.

  In case (c), $D$ has exactly one connected component of proper thinness $2$ and
  at least one other, all of the others of proper thinness $1$, so \cref{prop:pthin-2-1-multiple} gives $\pthin(H)=2$ if
  and only if $\mathit{stag}(Comp_i)$ holds, and otherwise
  $\pthin(H)=3$, because $2=\pthin(D)\le\pthin(H)\le\pthin(D)+1$ by \cref{thm:hereditary,thm:join}. When $\pthin(H)=2$, \cref{lem:staggered-construction,prop:pthin-2-1-multiple} give
  $\mathit{stag}(H)=\mathit{stag}(Comp_i)=\textsc{true}$, and
  \cref{lemma:optimal_boundary_solution} gives $\mathit{bnd}(H)=\mathit{bnd}(Comp_i)$.

  In case (d) we have $\pthin(D)=2$, and $D$ has exactly two connected
  components of proper thinness $2$, so \cref{prop:pthin-of-two-max-component} gives
  $\pthin(H)=2$ if and only if both of them admit an optimal boundary
  solution, and otherwise $\pthin(H)=3$ as in case (c). When
  $\pthin(H)=2$, we have $\mathit{stag}(H)=\textsc{false}$ by
  \cref{lem:two_components_no_staggered}, and $\mathit{bnd}(H)=\textsc{false}$ by \cref{lemma:optimal_boundary_solution},
  since $D$ has two connected components of proper thinness $2$. Case (e) is \cref{prop:pthin-of-three-or-more-max-component}.
\end{proof}

Item~2 of \cref{prop:dp} requires \(k\ge2\); by \cref{def:tree}, this holds at every extension node, which is why the tree of operations is normalized.

Evaluating \cref{prop:dp} bottom-up on the tree of operations gives the main algorithmic result of this section.

\begin{theorem}\label{theorem:linearAlgorithm}
    There exists a linear-time algorithm to determine whether the proper thinness of a trivially perfect graph is at most \(2\).
\end{theorem}
\begin{proof}
The algorithm evaluates \cref{prop:dp} on \(T(G)\). It first computes \(T(G)\) (\cref{thm:tree-linear}), and then traverses it bottom-up, computing \(\mathrm{st}(G_x)\) for every leaf and every extension node \(x\). To this end, every union node \(y\) passes to its parent three pieces of information: the maximum of the values \(p\) of its children, whether \(y\) has exactly two children and both of them are complete graphs, and the states of at most three of its children with \(p = 2\). Finally, the algorithm returns \(p(G_r)\) if the root \(r\) of \(T(G)\) is a leaf or an extension node, and the maximum computed at \(r\) if it is a union node. Computing \(T(G)\) takes \(O(n+m)\) time by
\cref{thm:tree-linear}; each of its nodes is then visited once, with constant work plus a constant amount per child,
so the traversal takes \(O(|T(G)|)=O(n)\) time, and the total is \(O(n+m)\).
\end{proof}

\section{Forbidden induced subgraphs of trivially perfect graphs}

\label{sec:forbidden}

In this section, we characterize trivially perfect graphs \(G\) with \(\pthin(G)\le 2\) in terms of forbidden induced subgraphs.

\begin{theorem}\label{thm:forbidden-minimal}
Let \(G\) be a trivially perfect graph. Then \(\pthin(G) \le 2\) if and only if \(G\) contains neither of the following graphs as an induced subgraph:
\begin{itemize}
    \item the \(\mathrm{claw}_2\) (\cref{fig:claw_2});
    \item the $\lotus$  (\cref{fig:lotus}).
\end{itemize}
Moreover, both graphs are minimal forbidden induced subgraphs for the class of trivially perfect graphs with proper thinness at most \(2\).
\end{theorem}

Its proof has two parts: the two obstructions have proper thinness \(3\) and are minimal, with the details in \ref{app:minimality}, and every trivially perfect graph with proper thinness at least \(3\) contains one of them as an induced subgraph. We begin by defining the two obstructions.

\begin{definition}\label{def:2claw}
The \emph{$\mathrm{claw}_2$} is the graph obtained from the disjoint union of three copies of \(K_{1,3}\), denoted \(3\cdot K_{1,3}\), by adding a single universal vertex (see \cref{fig:claw_2}).
\end{definition}

\begin{definition}
\label{def:lotus}

The \emph{$\lotus$} is the graph obtained from the disjoint union of two copies of \(K_{1,3}\) by successively adding a universal vertex, an isolated vertex, and then a second universal vertex (see \cref{fig:lotus}).
\end{definition}

\begin{figure}[H]
\centering
\begin{minipage}{0.48\textwidth}
 \centering
  \begin{tikzpicture}[scale=0.50,
    ue/.style={black, thick},
    ce/.style={black, thick},
    v/.style={circle, fill=black, inner sep=0pt, minimum size=7pt},
    ]
    \coordinate (u) at (0,0);
    \foreach \k/\t in {1/90, 2/210, 3/330}{
      \coordinate (c\k) at (\t:2.6);
      \coordinate (a\k) at ({\t-32}:4.6);
      \coordinate (b\k) at (\t:4.6);
      \coordinate (d\k) at ({\t+32}:4.6);
    }
    \foreach \k in {1,2,3}{
      \draw[ue] (u) -- (c\k);
      \draw[ue] (u) -- (a\k);
      \draw[ue] (u) -- (d\k);
      \draw[ue] (u) to[bend right=30] (b\k);
    }
    \foreach \k in {1,2,3}{
      \draw[ce] (c\k) -- (a\k);
      \draw[ce] (c\k) -- (b\k);
      \draw[ce] (c\k) -- (d\k);
    }
    \foreach \k in {1,2,3}{
      \node[v] at (c\k) {};
      \node[v] at (a\k) {};
      \node[v] at (b\k) {};
      \node[v] at (d\k) {};
    }
    \node[v] at (u) {};
  \end{tikzpicture}
  \caption{The $\mathrm{claw}_2$.}
  \label{fig:claw_2}
  \end{minipage}
\hfill
  \begin{minipage}{0.48\textwidth}
\centering
  \begin{tikzpicture}[scale=0.50,
    ue/.style={black, thick},
    ce/.style={black, thick},
    v/.style={circle, fill=black, inner sep=0pt, minimum size=7pt},
    ]
    \coordinate (u1) at (0,-0.4);
    \coordinate (u2) at (0,-2.2);
    \coordinate (w)  at (0,-4.0);
    \foreach \k/\s in {1/1, 2/-1}{
      \coordinate (c\k) at ({\s*(-3.2)}, 1.6);
      \coordinate (a\k) at ({\s*(-2.4)}, 3.0);
      \coordinate (b\k) at ({\s*(-3.5)}, 3.2);
      \coordinate (d\k) at ({\s*(-4.8)}, 1.3);
    }
    \foreach \k in {1,2}{
      \draw[ce] (c\k) -- (a\k);
      \draw[ce] (c\k) -- (b\k);
      \draw[ce] (c\k) -- (d\k);
    }
    \foreach \k in {1,2}{
      \foreach \x in {a,b,c,d}{ \draw[ue] (u1) -- (\x\k); }
    }
    \foreach \k in {1,2}{
      \foreach \x in {a,b,c,d}{ \draw[ue] (u2) -- (\x\k); }
    }
    \draw[ue] (u2) -- (u1);
    \draw[ue] (u2) -- (w);
    \foreach \k in {1,2}{
      \foreach \x in {a,b,c,d}{ \node[v] at (\x\k) {}; }
    }
    \node[v] at (u1) {};
    \node[v] at (u2) {};
    \node[v] at (w)  {};
  \end{tikzpicture}
  \caption{The \lotus{}.}
  \label{fig:lotus}
  \end{minipage}
\end{figure}

The following two lemmas give the first part; their proofs go through every graph obtained by deleting one vertex of each obstruction, and are deferred to \ref{app:minimality}.

\begin{lemma}
\label{lem:claw_2_minimal}
The graph $\mathrm{claw}_2$ has proper thinness $3$ and is a minimal forbidden induced subgraph for the class of graphs of proper thinness at most \(2\).
\end{lemma}

\begin{lemma}\label{lem:lotus_minimal}
The \lotus{} has proper thinness \(3\) and is a minimal forbidden induced subgraph for the class of graphs of proper thinness at most \(2\).
\end{lemma}

We now prove the second part.

\begin{lemma}[Forbidden induced subgraphs of proper thinness 3]\label{lem:forbidden-classification}
Let \(G\) be a trivially perfect graph with \(\pthin(G) \ge 3\). Then \(G\) contains the $\mathrm{claw}_2$ or the $\lotus$ as an induced subgraph.
\end{lemma}

\begin{proof}
Let \(H\) be an induced subgraph of \(G\) with \(\pthin(H) \ge 3\) that is minimal by number of vertices with this property, so that \(\pthin(H-v) \le 2\) for every \(v \in V(H)\). The graph \(H\) is trivially perfect, since this class is characterized by the forbidden induced subgraphs \(P_4\) and \(C_4\)~\cite{Golumbic1978}, and is hence hereditary. We show that \(H\), and hence \(G\), contains the $\mathrm{claw}_2$ or the $\lotus$ as an induced subgraph.

By \cref{thm:union} and the minimality of $H$, the graph $H$ is connected, so, being trivially perfect, it has a universal vertex \(u\), which is its only universal vertex by \cref{prop:many_universal_vertices} and the minimality of \(H\). Let \(H' = H - u\), again trivially perfect, with \(\pthin(H') \leq 2\) by the minimality of \(H\); moreover, \(H'\) is disconnected, as otherwise $H$ would have two universal vertices. Let \(\mathcal{C}_2\) be the family of connected components of \(H'\) with proper thinness \(2\), and let \(c_2 = |\mathcal{C}_2|\). Each component in \(\mathcal{C}_2\) contains an induced claw, since by \cref{lem:claw-free-pthin1} a claw-free trivially perfect graph has proper thinness \(1\).

For the cases \(c_2 = 2\) and \(c_2 = 1\), the following claim finds, inside a component without optimal boundary solutions, a universal vertex whose removal leaves two components of proper thinness~\(2\).

\begin{claim}\label{claim:no-boundary-two-components}
Let \(F\) be a connected trivially perfect graph with \(\pthin(F) = 2\) and no optimal boundary solution. Then \(F\) has a connected induced subgraph \(F'\) with a universal vertex \(u_1\) such that \(F' - u_1\) has two connected components \(D_1\) and \(D_2\) of proper thinness \(2\).
\end{claim}

\begin{proof}
Let \(F'\) be a connected induced subgraph of \(F\) with proper thinness \(2\) and no optimal boundary solution, minimal by number of vertices (\(F\) is a candidate). Being trivially perfect, \(F'\) has a universal vertex \(u_1\), and every component of \(F' - u_1\) has proper thinness at most \(2\) by \cref{thm:hereditary}. We apply \cref{lemma:optimal_boundary_solution} to the graph \(F' - u_1\) and the universal vertex \(u_1\). If no component of \(F' - u_1\) had proper thinness \(2\), then \(\pthin(F' - u_1) = 1\) by \cref{thm:union}, so \(F'\) would admit an optimal boundary solution, a contradiction. If exactly one component \(D\) had proper thinness \(2\), then \(D\) would have no optimal boundary solution, contradicting the minimality of \(F'\). Hence \(F' - u_1\) has two components \(D_1\) and \(D_2\) of proper thinness \(2\).
\end{proof}

\begin{itemize}
\item \textbf{$c_2 \ge 3$.}
  A claw in each of three components of proper thinness \(2\), together with \(u\), induces a $\mathrm{claw}_2$.

\item \textbf{$c_2 = 2$.}
  Let \(\CompOne\) and \(Comp_2\) be the two connected components of proper thinness \(2\) of $H'$. By \cref{prop:pthin-of-two-max-component}, one of them, say \(\CompOne\), lacks an optimal boundary solution. By \cref{claim:no-boundary-two-components}, applied to \(F = \CompOne\), there are a connected induced subgraph \(\CompOne'\) of \(\CompOne\) with a universal vertex \(u_1\) and two connected components \(D_1\) and \(D_2\) of \(\CompOne' - u_1\) of proper thinness \(2\). Each of \(D_1\) and \(D_2\) contains a claw by \cref{lem:claw-free-pthin1}, and so does \(Comp_2\). The three claws together with \(u\) induce a $\mathrm{claw}_2$ in \(H\) (\cref{def:2claw}).

\item \textbf{$c_2 = 1$.}
Let \(\CompOne\) be the unique component of \(H'\) with \(\pthin(\CompOne)=2\). All remaining components have proper thinness \(1\), and there is at least one of them because \(H'\) is disconnected; let \(w\) be a vertex of one of them. By \cref{prop:pthin-2-1-multiple}, \(\CompOne\) does not admit an optimal staggered solution, and hence, by \cref{boundary_implies_staggered}, which applies since \(\pthin(\CompOne)=2\), no optimal boundary solution either. By \cref{claim:no-boundary-two-components}, applied to \(F = \CompOne\), there are a connected induced subgraph \(\CompOne'\) of \(\CompOne\) with a universal vertex \(u_1\) and two connected components \(D_1\) and \(D_2\) of \(\CompOne' - u_1\) of proper thinness \(2\). Each of \(D_1\) and \(D_2\) contains a claw by \cref{lem:claw-free-pthin1}, and the two claws together with \(u_1\), \(w\) and \(u\) induce a $\lotus$ in \(H\) (\cref{def:lotus}).

\item \textbf{$c_2 = 0$.}
Then \(\pthin(H') = 1\) by \cref{thm:union}, so \(\pthin(H) \le 2\) by \cref{thm:join}, a contradiction.
\end{itemize}

Thus, \(H\), and hence \(G\), contains the $\mathrm{claw}_2$ or the $\lotus$ as an induced subgraph.
\end{proof}

\medskip
\noindent Putting \cref{lem:forbidden-classification,lem:claw_2_minimal,lem:lotus_minimal} together, we obtain \cref{thm:forbidden-minimal}.

\begin{proof}[Proof of \cref{thm:forbidden-minimal}]
Both obstructions have proper thinness \(3\) (\cref{lem:claw_2_minimal} and \cref{lem:lotus_minimal}), so a graph \(G\) containing either of them as an induced subgraph has \(\pthin(G) \ge 3\) by \cref{thm:hereditary}; hence \(\pthin(G) \le 2\) implies that \(G\) contains neither, which is the forward direction. The reverse direction is \cref{lem:forbidden-classification}. Both obstructions are trivially perfect, since they are built from single vertices by disjoint unions and universal extensions (\cref{def:triviallyPerfect-recursive}), and their minimality follows from \cref{lem:claw_2_minimal} and \cref{lem:lotus_minimal}.
\end{proof}

\section{Conclusions and future work}
\label{sec:conclusions}

\paragraph{Proper thinness greater than \texorpdfstring{\(2\)}{2}}
Our algorithm does not determine the value of the parameter when it is at least~\(3\).

\begin{open}\label{open:pthin-tp}
Is there a polynomial-time algorithm that computes \(\pthin(G)\) for every trivially perfect graph \(G\)? Or at least, for a fixed \(k \ge 3\), one that decides whether \(\pthin(G) \le k\) for a trivially perfect graph \(G\)?
\end{open}

The approach of \cref{sec:triviallyperfect} suggests a strategy: for \(k\) classes, one would need an analogous description of how much room an optimal solution leaves for the next operation, and to show that it can be computed from the same description of the graphs being combined. For \(k = 2\) the recursion closes after a single step, since the staggered and the boundary conditions of \(\Gplusu\) are determined by the same two conditions on the components of \(G\). A first attempt for \(k = 3\) suggests that finer properties are needed, and that the chain of properties becomes much longer before it closes.

\paragraph{Forbidden induced subgraphs for larger values}
A positive answer to the following problem for some fixed \(k\) would also give a positive answer to the second question of \cref{open:pthin-tp} for that \(k\), since deciding whether a graph contains a fixed graph as an induced subgraph can be done in polynomial time.

\begin{open}\label{open:forbidden-k}
For a fixed \(k \ge 3\), is the family of minimal forbidden induced subgraphs for the class of trivially perfect graphs with proper thinness at most \(k\) finite? If so, can it be described explicitly?
\end{open}

\paragraph{A certifying algorithm}
The proofs on which our algorithm relies are constructive. It would be interesting to extend the algorithm so that it also returns a strongly consistent solution with two classes when \(\pthin(G) \le 2\), and an induced \(\mathrm{claw}_2\) or \lotus{} when \(\pthin(G) \ge 3\), while keeping the linear running time.

\paragraph{Larger classes}
Since the thinness of a cograph can be computed in polynomial time~\cite{B-D-thinness}, the class of cographs is the natural next step.

\begin{open}\label{open:cographs}
Is there a polynomial-time algorithm that computes the proper thinness of a cograph?
\end{open}

Our analysis does not extend directly to cographs: a universal vertex occupies a single position of the ordering and belongs to a single class, whereas in a join of two arbitrary graphs the vertices of both sides may be interleaved. Interval graphs~\cite{Brandstadt-book} are another natural candidate. Finally, the complexity of computing the proper thinness of a tree is also open, although the thinness of trees can be computed in polynomial time~\cite{Agus-Eric-DAM25} and the trees of proper thinness at most~\(2\) have been characterized~\cite{B-M-P-trees-pthin2}.

\section*{Declaration of generative AI and AI-assisted technologies in the manuscript preparation process}

During the preparation of this work the authors used Claude Opus 5.5 (by Anthropic) to improve the language and readability of the manuscript and to assist with LaTeX typesetting. After using this tool, the authors reviewed and edited the content as needed and take full responsibility for the content of the published article.

\section*{Acknowledgments}

\begin{sloppypar}
This work was partially supported by UBACyT (20020220300079BA and 20020190100126BA) and CONICET (PIP 11220200100084CO and PIP 11220250101365CO). B.~Rivero was supported by two Becas de Iniciaci\'on a la Investigaci\'on en Ciencias de la Computaci\'on (BIICC), in 2025 and 2026, and by a scholarship from LambdaClass.
\end{sloppypar}

\bibliographystyle{elsarticle-num-names}
\bibliography{paper}

\appendix
\renewcommand{\thetheorem}{A.\arabic{theorem}}
\renewcommand{\thelemma}{A.\arabic{theorem}}

\section{Proper thinness and minimality of the obstructions}
\label{app:minimality}

In this appendix we prove \cref{lem:claw_2_minimal,lem:lotus_minimal}. Instead of exhibiting a strongly consistent solution of each graph obtained by deleting a vertex, we derive all the cases from the results of \cref{sec:triviallyperfect}, through \cref{lem:claw-pieces} below.
The results of \cref{sec:triviallyperfect} apply because all the graphs considered in the following proofs are built from single vertices by disjoint unions and universal extensions, and are hence trivially perfect (\cref{def:triviallyPerfect-recursive}). We first obtain the two families of graphs on which most of the following proofs rely.

\begin{lemma}
\label{lem:claw-pieces}
Let \(H\) be a graph that is either empty or trivially perfect with proper thinness 1.
\begin{enumerate}
    \item The graph obtained by adding a universal vertex to the disjoint union of two claws and \(H\) has proper thinness \(2\).
    \item If \(H\) is not empty, the graph \(D\) obtained by adding a universal vertex to the disjoint union of a claw and \(H\) has proper thinness \(2\) and admits an optimal staggered solution.
\end{enumerate}
\end{lemma}
\begin{proof}
The claw, of proper thinness \(2\) (\cref{obs:claw-pthin2}), is obtained by adding a universal vertex to three isolated vertices, whose disjoint union has proper thinness \(1\), so it admits an optimal boundary solution by \cref{lemma:optimal_boundary_solution}, and hence an optimal staggered solution by \cref{boundary_implies_staggered}.

\smallskip
\noindent\textbf{Item 1.} Let \(G\) be the disjoint union of two claws and \(H\), so that the graph of item~1 is \(\Gplusu\). Its connected components are the two claws, of proper thinness \(2\), and the connected components of \(H\), if any, of proper thinness \(1\) by \cref{thm:hereditary}, so \(\pthin(G) = 2\) by \cref{thm:union}. Both claws admit an optimal boundary solution, so \cref{prop:pthin-of-two-max-component} gives \(\pthin(\Gplusu) = 2\).

\smallskip
\noindent\textbf{Item 2.} Let \(G\) be the disjoint union of the claw and \(H\), so that \(D = \Gplusu\). The connected components of \(G\) are the claw, of proper thinness \(2\) (\cref{obs:claw-pthin2}), which admits an optimal staggered solution, and those of \(H\), of proper thinness \(1\) by \cref{thm:hereditary}. Hence, by \cref{lem:staggered-construction}, \(D\) admits a staggered solution using two classes, which is optimal since \(\pthin(D) \ge 2\) by \cref{obs:claw-pthin2,thm:hereditary}.
\end{proof}

\begin{proof}[Proof of \cref{lem:claw_2_minimal}]
By \cref{obs:claw-pthin2}, $\pthin(\mathrm{claw}_2)=3$. For minimality, let $v$ be any vertex of $\mathrm{claw}_2$; we show that $\pthin(\mathrm{claw}_2-v)\leq 2$. Any two vertices of $\mathrm{claw}_2$ with the same degree are exchanged by an automorphism, so there are three cases.
\begin{itemize}
    \item $v$ is the universal vertex, with \(d(v) = 12\). Then $\mathrm{claw}_2 - v$ is the disjoint union of three claws, which has proper thinness \(2\) by \cref{thm:union,obs:claw-pthin2}.
    \item $v$ is the center of one of the claws, with \(d(v) = 4\). Then $\mathrm{claw}_2 - v$ is obtained by adding a universal vertex to the disjoint union of the two other claws and three isolated vertices, so it has proper thinness \(2\) by item~1 of \cref{lem:claw-pieces}, with the graph \(H = 3K_1\).
    \item $v$ is a leaf of one of the claws, with \(d(v) = 2\). Then $\mathrm{claw}_2 - v$ is obtained by adding a universal vertex to the disjoint union of the two other claws and \(K_{1,2}\), so it has proper thinness \(2\) by item~1 of \cref{lem:claw-pieces}, with the graph \(H = K_{1,2}\).
\end{itemize}

By the cases above and \cref{thm:hereditary}, every proper induced subgraph of $\mathrm{claw}_2$ has proper thinness at most $2$, so $\mathrm{claw}_2$ is a minimal forbidden induced subgraph for the class of graphs of proper thinness at most \(2\).
\end{proof}

\begin{proof}[Proof of \cref{lem:lotus_minimal}]
Let \(G'\) be the graph obtained by adding a universal vertex \(u_1\) to the disjoint union of two claws, and let \(G\) be the disjoint union of \(G'\) and a new vertex \(w\). Then the \lotus{} is the graph \(\Gplusu\) with \(u = u_2\). By item~1 of \cref{lem:claw-pieces}, with \(H\) empty, \(\pthin(G') = 2\). Since \(G'\) is obtained by adding a universal vertex to the disjoint union of two claws, which has proper thinness \(2\) and two connected components of proper thinness \(2\), and \(\pthin(G')=2\), \cref{lem:two_components_no_staggered} shows that \(G'\) does not admit an optimal staggered solution. The graph \(G\) has proper thinness \(2\) by \cref{thm:union}, and \(G'\) is its only connected component of proper thinness \(2\). Since \(G'\) does not admit an optimal staggered solution, the implication \(1 \Rightarrow 2\) of \cref{prop:pthin-2-1-multiple} shows that \(\pthin(\Gplusu) \neq 2\). Moreover, \(\pthin(\Gplusu) \geq \pthin(G') = 2\) by \cref{thm:hereditary}, and \(\pthin(\Gplusu) \leq \pthin(G) + 1 = 3\) by \cref{thm:join}. Hence \(\pthin(\lotus) = 3\).

For minimality, let $v$ be any vertex of $\lotus$; we show that $\pthin(\lotus-v)\leq 2$. Any two vertices of $\lotus$ with the same degree are exchanged by an automorphism, so there are five cases.
\begin{itemize}
    \item $v = u_2$, with \(d(v) = 10\). Then \(\lotus - v = G' \cup \{w\}\), which has proper thinness \(2\) by \cref{thm:union}.
    \item $v = u_1$, with \(d(v) = 9\). Then \(\lotus - v\) is obtained by adding the universal vertex \(u_2\) to the disjoint union of the two claws and \(w\), so it has proper thinness \(2\) by item~1 of \cref{lem:claw-pieces}, with the graph \(H\) induced by the single vertex $w$.
    \item $v = w$, with \(d(v) = 1\). Then \(\lotus - v\) is obtained by adding the universal vertex \(u_2\) to the connected trivially perfect graph \(G'\), so \(\pthin(\lotus - v) = \pthin(G') = 2\) by \cref{prop:many_universal_vertices}.
    \item $v$ is the center of one of the claws, with \(d(v) = 5\), or a leaf of one of the claws, with \(d(v) = 3\). Let \(H\) be what remains of that claw, namely the graph \(3K_1\) or \(K_{1,2}\), respectively, and let \(D = \lotus - v - u_2 - w\), which is obtained by adding the universal vertex \(u_1\) to the disjoint union of the other claw and \(H\). By item~2 of \cref{lem:claw-pieces}, \(D\) has proper thinness \(2\) and admits an optimal staggered solution. The graph \(\lotus - v\) is obtained by adding the universal vertex \(u_2\) to the disjoint union of \(D\) and \(w\), whose only connected component of proper thinness \(2\) is \(D\), so it has proper thinness \(2\) by the implication \(2 \Rightarrow 1\) of \cref{prop:pthin-2-1-multiple}.
\end{itemize}

By the cases above and \cref{thm:hereditary}, every proper induced subgraph of the \lotus{} has proper thinness at most $2$, so the \lotus{} is a minimal forbidden induced subgraph for the class of graphs of proper thinness at most \(2\).
\end{proof}

\end{document}